\documentclass[sigconf]{acmart}
\AtBeginDocument{%
  }

\usepackage{xcolor}
\usepackage{fontawesome} 
\usepackage[capitalize, noabbrev]{cleveref}
\usepackage{siunitx}
\usepackage{amsmath}      %
\usepackage{booktabs}     %
\usepackage{tabularx}
\usepackage{makecell}
\usepackage{enumitem}

\setcopyright{acmlicensed}
\copyrightyear{2018}
\acmYear{2018}
\acmDOI{XXXXXXX.XXXXXXX}
\acmConference[Conference acronym 'XX]{Make sure to enter the correct
  conference title from your rights confirmation email}{June 03--05,
  2018}{Woodstock, NY}
\acmISBN{978-1-4503-XXXX-X/2018/06}

\newcommand\revision[1]{\textcolor{black}{#1}}

\newcommand{\pct}[1]{\SI{#1}{\percent}}
\newcommand{\nMain}[0]{184} %
\newcommand{\nRec}[0]{200}
\newcommand{\nExcl}[0]{16}
\newcommand{\noAI}[0]{\textit{noAI}}
\newcommand{\withAI}[0]{\textit{withAI}}

\copyrightyear{2026}
\acmYear{2026}
\setcopyright{cc}
\setcctype{by}
\acmConference[CHI '26]{Proceedings of the 2026 CHI Conference on Human Factors in Computing Systems}{April 13--17, 2026}{Barcelona, Spain}
\acmBooktitle{Proceedings of the 2026 CHI Conference on Human Factors in Computing Systems (CHI '26), April 13--17, 2026, Barcelona, Spain}
\acmPrice{}
\acmDOI{10.1145/3772318.3791494}
\acmISBN{979-8-4007-2278-3/2026/04}

\begin{document}

\title{The AI Memory Gap: Users Misremember What They Created With AI or Without}

\author{Tim Zindulka}
\email{tim.zindulka@uni-bayreuth.de}
\orcid{0009-0009-1972-351X}
\affiliation{%
  \department{Department of Computer Science}
  \institution{University of Bayreuth}
  \city{Bayreuth}
  \country{Germany}
}

\author{Sven Goller}
\email{sven.goller@uni-bayreuth.de}
\orcid{0000-0001-5263-5372}
\affiliation{%
  \department{Department of Computer Science}
  \institution{University of Bayreuth}
  \city{Bayreuth}
  \country{Germany}
}

\author{Daniela Fernandes}
\email{daniela.dasilvafernandes@aalto.fi}
\orcid{0009-0006-1332-7485}
\affiliation{%
  \institution{Aalto University}
  \city{Espoo}
  \country{Finland}
}

\author{Robin Welsch}
\email{robin.welsch@aalto.fi}
\orcid{0000-0002-7255-7890}
\affiliation{%
  \institution{Aalto University}
  \city{Espoo}
  \country{Finland}
}

\author{Daniel Buschek}
\email{daniel.buschek@uni-bayreuth.de}
\orcid{0000-0002-0013-715X}
\affiliation{%
  \department{Department of Computer Science}
  \institution{University of Bayreuth}
  \city{Bayreuth}
  \country{Germany}
}

\renewcommand{\shortauthors}{Zindulka et al.}

\begin{abstract}
As large language models (LLMs) become embedded in interactive text generation, disclosure of AI as a source depends on people remembering which ideas or texts came from themselves and which were created with AI.
We investigate how accurately people remember the source of content when using AI. In a pre-registered experiment, \nMain{} participants generated and elaborated on ideas both unaided and with an LLM-based chatbot. One week later, they were asked to identify the source (noAI vs withAI) of these ideas and texts.
Our findings reveal a significant gap in memory: After AI use, the odds of correct attribution dropped, with the steepest decline in mixed human-AI workflows, where either the idea or elaboration was created with AI. We validated our results using a computational model of source memory.
Discussing broader implications, we highlight the importance of considering source confusion in the design and use of interactive text generation technologies.

\end{abstract}

\ccsdesc[500]{Human-centered computing~Empirical studies in HCI}
\ccsdesc[500]{Computing methodologies~Natural language generation}

\keywords{Ideation, Writing, AI, LLM, Source Memory}

\begin{teaserfigure}
\centering
\includegraphics[width=.93\textwidth]{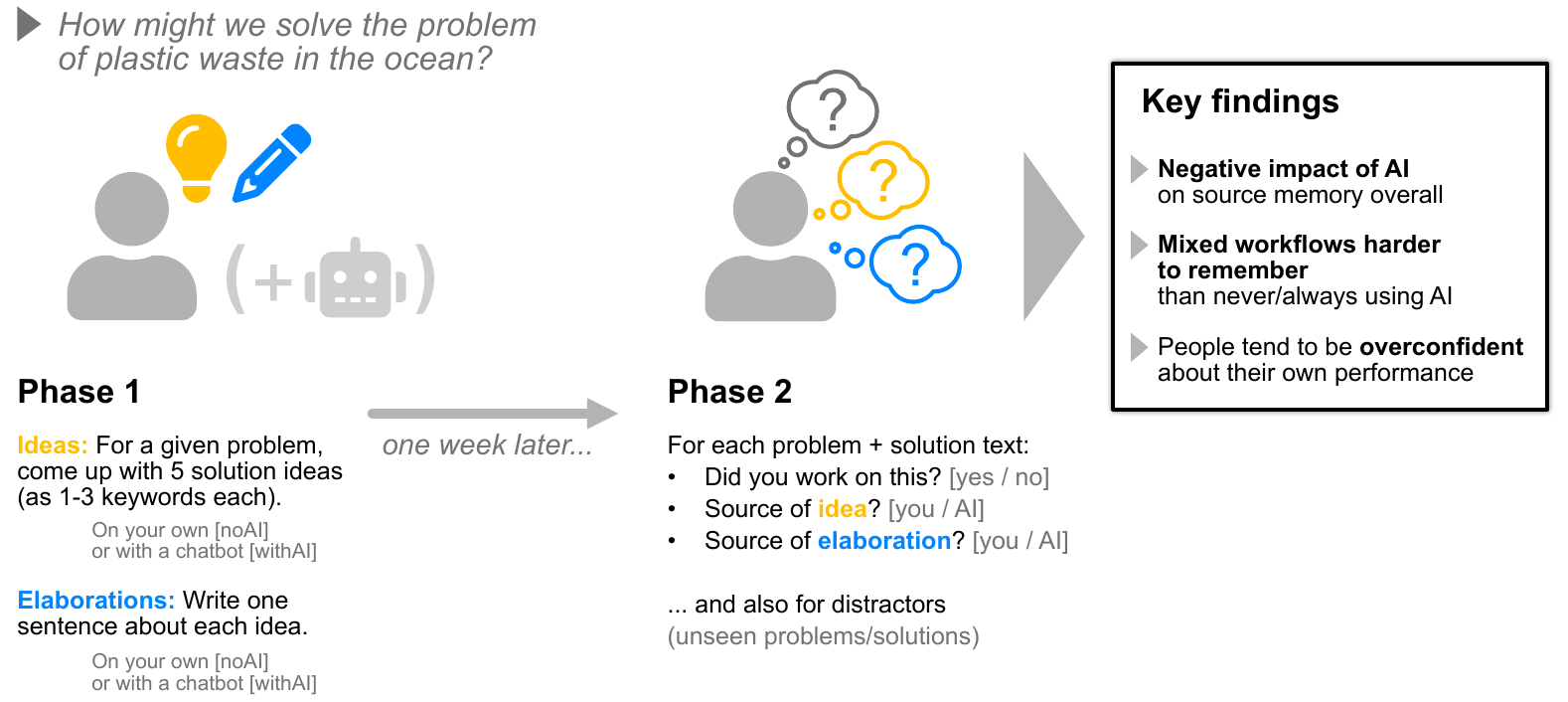}
\caption{Overview of our experiment and findings: In Phase 1 (left) participants were presented with problems and asked to come up with five ideas each (as 1-3 keywords per idea), with or without assistance by a chatbot (alternating across problems). %
They then wrote on these ideas (1 sentence per idea), again with or without AI. After a week, Phase 2 (centre) presented the problems and solutions again, asking participants if they worked on this (item memory), and if so, what the source was for both idea and elaboration (source memory). We also showed distractors (unseen problems/solutions). The key findings (right) show that AI had a negative impact, mixed workflows were particularly difficult, and people tended to be overconfident in their own performance.}
\Description{Illustration showing the study flow overall: Left part shows Phase 1 with a user and chatbot working on ideas and elaborations for the example problem ``How might we solve the problem of plastic waste in the ocean?'' Text says that participants have to come up with five ideas in 1-3 keywords and write elaborations in one sentence each. An arrow with annotation (``One week later'') points to a second illustration of a user with question-marked though bubbles. Text says that in this Phase 2, participants were asked: Did you work on this? Source of idea? Source of elaboration? Further text says that they were also asked this for unseen items (so-called distractors). Following another arrow to the right is a box with the title ``Key findings'' and three bullets: Negative impact of AI on source memory overall; mixed workflows harder to remember than never/always using AI; and people tend to be overconfident about their own performance.}
\label{fig:teaser}
\end{teaserfigure}

\maketitle

\section{Introduction}
Do you remember which ideas and sentences you authored yourself and which you created together with AI?

Co-creation with Large Language Models (LLMs) is rapidly evolving, as many people integrate AI into their daily workflows for a wide range of tasks~\cite{Mysore2025behaviors},
with a growing body of research designing LLM-based systems for writing~\cite{Lee2024designspace, Benharrak24persona, Hoque2024hallmark}, ideation~\cite{Shin2023collabIdeationWorkshop, Suh24luminate, Shaer2024brainwriting}, developing concepts~\cite{Han2024teamsai, Wang2025aideation}, and solving complex problems~\cite{he2025finegrainedappropriatereliancehumanai}, which comes with metacognitive demands~\cite{Tankelevitch2024metacognitivedemands}.
However, despite the growing reliance on LLMs, little is known about people's ability to accurately recall their own contributions in content co-created with AI.

As research shows, users generally perform at near-chance levels when trying to identify AI‑generated text~\cite{clark2021thatshumangoldevaluating, Jakesch_heuristics_2023, boutadjine2025human}, particularly non-regular users~\cite{russell2025peoplefrequentlyusechatgpt}.
Work with essay writing found that LLM users recalled less from their own texts and showed weaker EEG engagement than unaided or search users, though in a small sample~\cite{kosmyna2025brainonchatgpt}.
However, there is no systematic study on how writers remember the \textit{sources} of ideas and texts after working with AI.

At the same time, there is growing concern around authorship claims in AI-generated content~\cite{credit_he2025, Kyi25_authorshipgov}, with evidence that users sometimes take credit for work largely or entirely produced by LLMs~\cite{draxler24_ghostwriter}.

This issue matters not only for questions of intellectual ownership, but also because AI can shape cognitive processes.
For instance, prior research shows that \revision{AI-edited images and videos can induce false recollections~\cite{Pataranutaporn2025synthetic_memories}. Similarly, co-creating or revising text with AI may blur the line between own experience and AI-generated modifications, potentially influencing source memory. 
Furthermore,} \citet{jakesch23_opinionated} found that LLMs can influence users' opinions, which suggests they might also affect how people remember their own ideas and contributions.

Decades of cognitive psychology indicate that when two sources produce highly similar linguistic output, especially within a single writing session, people rely on heuristic cues (e.g., ease of retrieval, writing fluency) rather than firm recollection to attribute authorship~\cite{Johnson1993SourceMonitoring}. 
LLM co‑writing blurs those cues further because AI suggestions are (1) context‑specific, (2) immediately editable, and (3) increasingly integrated into the user's document editing tools, making them potentially feel \emph{self-generated}. 

This new form of source confusion between self and AI could have important implications for perceived authorship, learning, accountability, and trust in both self and system.
It may shape how users reflect on and take ownership of their work, particularly in educational, professional, or creative contexts.
For instance, users could think less of their own expertise by discovering an AI-generated mistake and misremembering it as their own. 
Or they might treat a document as more reliable than warranted if they misremember AI-generated content as their own. 
Conversely, they may lose deserved ``credit'' in external evaluations by mistakenly disclosing an own idea as AI-generated.

\revision{As AI tools and their capabilities evolve, the need to track and disclose AI involvement will remain a recurring challenge. This prompts institutions to frequently update their policies, requiring post hoc declarations of when and where AI was used.}
\revision{For example, academic journals may mandate disclosure of AI assistance for manuscripts drafted before such rules were in place (e.g., Elsevier's update on GenAI use\footnote{\url{https://www.elsevier.com/about/policies-and-standards/generative-ai-policies-for-journals}}), and in the legal domain, courts may require practitioners to specify which parts of a document were generated by AI, sometimes months after the original filing (cf. ``Mata v. Avianca, Inc.''~\cite{lyon2023chatgptsanctions}).}

\revision{Despite these issues,} we currently lack the fundamental understanding necessary to make informed design choices in this context.
In particular, it remains unknown how accurately and confidently people can recall what they created themselves versus what was suggested by AI, and whether this depends on the AI's role (e.g., idea generation, text composition, or both).
\revision{Thus, our study focuses on ideation and elaboration as fundamental elements of writing workflows. Separating these stages also allows us to examine how using AI at different points in the writing process affects people's memory.}
This motivates our research questions:
\begin{itemize}
    \item \textbf{RQ1:} How does working with AI on a problem affect people's \textit{ability and confidence in remembering} (1) the solution, (2) the source of the idea, and (3) the source of its elaboration?
    \item \textbf{RQ2:} How do mixed human-AI \textit{workflows} (different sources for ideas and elaborations) compare to consistent workflows in terms of memory accuracy and confidence in source attribution?
    \item \textbf{RQ3:} To what extent are people able to \textit{judge their own ability to remember} whether an idea or elaboration originated from AI vs. themselves?
\end{itemize}

To address these questions, we conducted a within-subjects study with two phases (\cref{fig:teaser}).
In Phase 1, participants (N=\nMain{}) completed ideation and writing tasks either alone (\noAI) or with an LLM-based chatbot (\withAI). 
After a one-week delay (Phase 2), they were asked to recall the origin of each idea and elaboration sentence they had produced, specifying whether they had created it themselves or with AI support, and rating their confidence in each decision. 
To probe memory accuracy more rigorously, we also introduced content not previously seen by participants (``distractors'').

Our results reveal an ``AI Memory Gap'', where any AI involvement impaired memory, specifically source memory, and most strongly in mixed human-AI workflows.
The odds of correctly attributing the source of ideas were 95\% lower when the idea originated from AI but was elaborated by the human, and 86\% lower when the idea was human but the elaboration was generated by AI, compared to the all-human baseline.
When participants consistently used AI in both steps, the interaction partly offset these losses.
The same pattern holds true for elaboration source attribution, where using AI for the idea alone reduced the odds of correct attribution by 65\%, and using AI for elaboration alone reduced them by 63\%. When both idea and elaboration came from AI, the interaction again partly compensated for the losses, but still remained below the all-human baseline.
Confidence mirrored these effects: Participants were most confident in the all-human workflow and less confident with AI involvement. Yet, they systematically overestimated their performance, with self-rated accuracy exceeding actual performance by 12\% for ideas, and 6\% for elaborations.

These findings show that co-creating with LLMs can systematically impact source memory and attribution, and that the role of AI within the workflow matters in ways that open avenues for UI and interaction design.

Discussing broader implications, we highlight the importance of considering source confusion in the design of interactive text generation technologies.

\section{Related Work}
\subsection{Ideation and Writing Assistants}

We investigate source memory for both finding ideas and expressing them in text. Thus, here we relate our work to research on ideation with AI and writing with AI.

\subsubsection{Ideation with AI}

Recent related work addresses HCI aspects of integrating AI into ideation processes: For example, \citet{Shin2023collabIdeationWorkshop} organised a CHI'23 workshop on collaborative ideation with AI and \citet{Han2024teamsai} investigated how pairs of students prompt ChatGPT and Midjourney to create stage design ideas. Similarly, \citet{MahdaviGoloujeh2024isitai} investigated users' ``prompting journeys'' for text-to-image generation via interviews, extracting common structures and challenges. Their title asks ``Is it AI or is it me?'' as a reflection on creatives' perception but not on remembering their creative actions.

Further work examined creatively stimulating effects of ``AI errors'', such as mislabelled object recognitions~\cite{van_der_burg2022ceci}. \citet{Liu2024aierror} investigate such errors specifically for ideation, to promote human creativity through serendipity and association. 

Other work reports controlled experiments on related (negative) effects: 
\citet{Guo2024alignment} studied the impact of value alignment between user and chatbot in a brainstorming task (i.e. both ``pro'' or ``con'' on a topic -- or in opposition). They found no difference in idea quality, but the generated ideas reflected the AI's values and using AI reduced the feeling of ownership.
Related, \citet{Wadinambiarachchi2024fixation} found that novice designers using image generators had higher design fixation (i.e. keeping close to features of an initial example). They also produced fewer, less varied, and less original ideas with AI.

In contrast, avoiding fixation is a design goal in other work: \textit{CreativeConnect} by \citet{Choi2024creativeconnect} helps graphic designers expand their space of ideas by extracting inspiring keywords from reference images on a moodboard.
Similarly, \textit{AIdeation} by \citet{Wang2025aideation} supports idea exploration with image generation and keyword extraction in an iterative ideation loop. And \textit{Luminate} by~\citet{Suh24luminate} plots systematically varied alternatives on a 2D space of ideas.

While the above systems have different roles for what the user and AI add to the process and UI, the \textit{IdeationWeb} system by \citet{Shen2025ideationweb} used colour-coded nodes to mark otherwise similar representations of human and AI ideas, namely nodes in a mind map-like diagram. In contrast, we examine how people remember the source of ideas themselves.

In summary, there is high interest in applying AI in ideation tasks, with research on both constructive aspects (e.g. tool design) and assessing impact. Adding to this literature, we examine the new aspect of how people later remember the source of ideas when using AI in ideation. Specifically, our task involves both coming up with ideas and writing short descriptions, leading us to a review of writing with AI next.

\subsubsection{Design and perception of writing with AI}
The recent survey by~\citet{Lee2024designspace} analysed 115 ``intelligent'' writing tools from papers at HCI and NLP venues, charting a design space with five high-level dimensions. These design choices may impact how people remember their interactions and results. For example, their dimension of ``Interaction'' includes aspects such as UI layout (separated or integrated AI), interaction metaphor (e.g. tool or agent), and interface paradigm (e.g. chatbot or text editor). 
\revision{For our study, we decided to replicate a common AI-assisted writing workflow: using an AI chatbot alongside a text editor. To keep both visible without switching apps/tabs, we embedded the chatbot into a sidebar within our UI, similar to placing a text editor (e.g., Word, Docs) and AI tool (e.g., ChatGPT, Claude) side by side.}

Moreover, this survey~\cite{Lee2024designspace} extracted aspects of the user's long-term perception and relationship to the system (e.g. agency, ownership, integrity, trust, transparency). Memory was not discussed, reflecting the gap in the literature, although remembering what was created by AI or the user seems highly relevant for long-term perception. For instance, perceived ownership may be impacted by this, as well as integrity -- for example, awareness of (accidental) plagiarism introduced by AI-generated content. 

One notable exception is the \textit{HaLLMark} system by \citet{Hoque2024hallmark}, which is explicitly designed for writers to keep a history of prompts and AI contributions throughout their writing process. As with \textit{IdeationWeb}~\cite{Shen2025ideationweb} above, this tracks AI use in the UI. %

\revision{Finally, earlier work on computer-supported collaborative work (CSCW) tracked and visualised \textit{human} contributions (e.g.~\cite{Bergstrom2007clock, Shi2023traces}). While some solutions for detecting and tracking AI use exist in research and products (e.g.~\cite{Gehrmann2019gltr, Hoque2024hallmark, iAWriter2025website}), our findings motivate revisiting the CSCW design space more broadly -- now for tracking AI. Some adaptations might be necessary: In contrast to human-human collaboration, there is no other person that might keep own notes or remember their contribution independent of what is tracked by the system, and people tend to give less credit to AI than a human writer~\cite{draxler24_ghostwriter}. Overall, we see our contribution as complementary to system/design work, in that we quantify an unaided human memory baseline. For example, \textit{HaLLMark}~\cite{Hoque2024hallmark} could not relate to memory at all in its design rationale, and framed it around accountability, ownership, and external policies. Our findings provide an important ``internal'' addition to this set: imperfect human memory of interaction with AI.} %

\subsection{Source Memory}
\revision{After relating our work to the literature on writing with AI, we now introduce the concept of source memory to contextualise these insights within established concepts from memory research.}

Source memory refers to the processes that enable people to remember \emph{where}, \emph{when}, and \emph{how} an item was acquired, in contrast to \emph{item memory}, which concerns only whether the item was encountered at all \cite{Johnson1993SourceMonitoring, mitchell2009source}.  The dominant theoretical account is the \textit{Source Monitoring Framework} (SMF), which posits that people attribute mental contents to a source by evaluating qualitative features of the memory trace -- such as perceptual details, spatio-temporal context, associated affect, and records of the cognitive operations that generated the content \cite{mitchell2009source}. When these features are insufficiently distinctive, individuals rely on heuristic cues such as familiarity or processing fluency, increasing the likelihood of misattribution.

Items that people \emph{generate} themselves are typically recognised better than items they merely read or hear, a robust phenomenon known as the \textit{generation effect} \cite{geghman2004generation}.  Paradoxically, generation can \textit{impair} source memory under conditions of high similarity between self- and other-produced material.  For instance,  \citet{landau1997monitoring} showed that when participants alternated between their own word puzzle solutions and copying sentences provided by an experimenter, they later misattributed large proportions of the material. Comparable effects -- unknowingly claiming someone else's ideas -- have been reported for collaborative recall and brainstorming tasks \cite{brown1989cryptomnesia,marsh1997contributions}, but not yet for ``collaboration'' with AI.  

Digital environments introduce additional challenges for source monitoring.  %
For instance, real-time collaboration and ubiquitous copy-paste affordances might blur perceptual and contextual cues that would otherwise separate one's own contributions from external input. Code repository studies of end-user programmers, for example, document frequent misattribution of copied code snippets to self-authorship \cite{baltes2019usage}. In language generation interfaces, suggestions are inserted directly into the working document; users can accept, reject, or overwrite them at the keystroke level, intertwining human and machine text in ways that make the final product perceptually homogeneous (e.g. see \cite{Buschek2021suggestions, Lee2022coauthor}). Under the SMF, such homogeneity should reduce cue distinctiveness and therefore impair source discrimination -- a prediction supported by recent empirical work on AI-assisted writing \cite{draxler24_ghostwriter} and on detecting AI-generated texts \cite{boutadjine2025human}.

Confidence in source decisions is only weakly correlated with accuracy (e.g. see \cite{horry2014confidence}). People consistently overestimate their ability to identify the origin of text, especially when they believe that stylistic idiosyncrasies are obvious cues such as in the case of AI writing \cite{fleckenstein2024teachers}. Metacognitive overconfidence is exacerbated by \textit{fluency} -- the ease with which information is processed -- as fluent processing is misinterpreted as evidence of prior self-generation \cite{huang2021examining}.  LLMs explicitly optimise for fluency, suggesting that AI collaboration may produce particularly compelling illusions of authorship.

Overall, despite a rich literature on source monitoring in laboratory tasks, little is known about source memory for content created in \emph{interactive human-AI} settings.

\section{Method}
\begin{figure*}
    \centering
    \includegraphics[width=\linewidth]{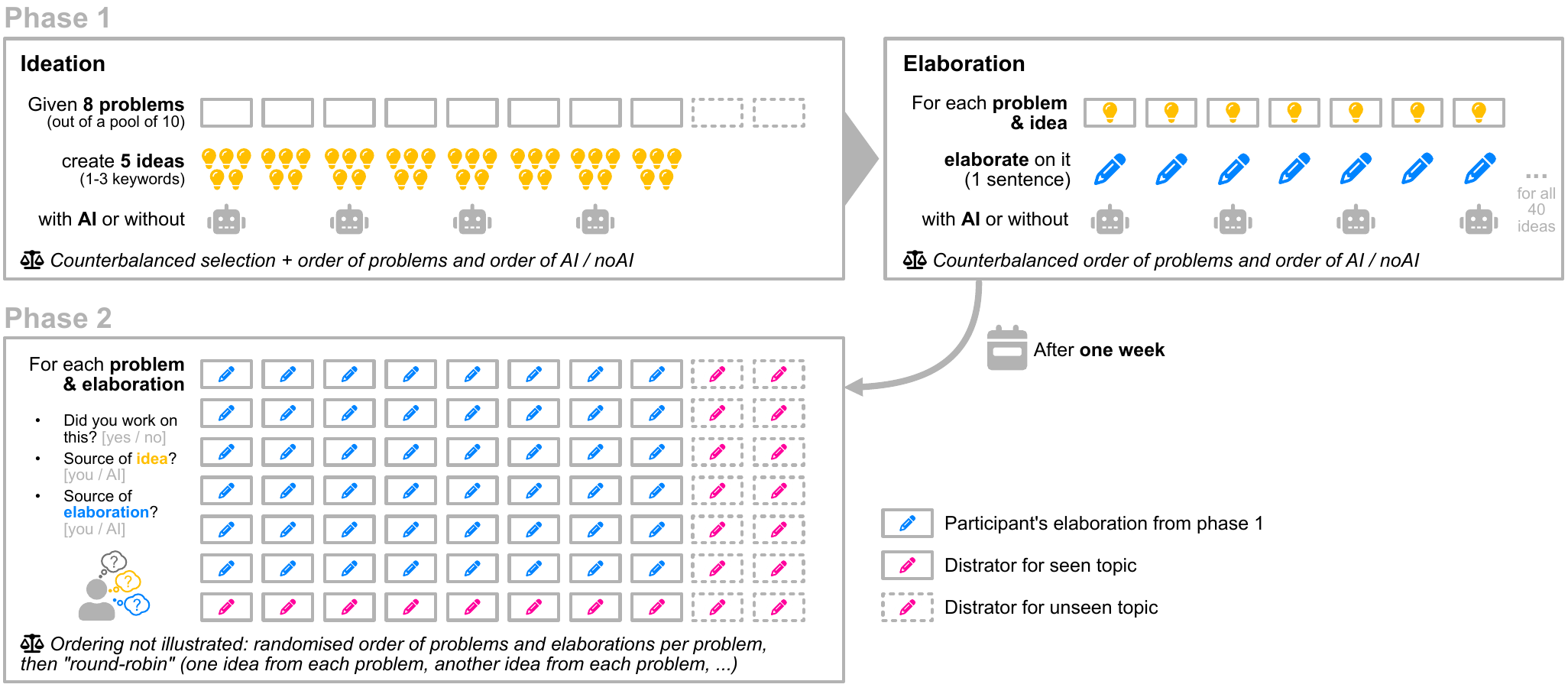}
    \caption{Study design and procedure. The experiment consisted of two phases separated by one week. In Phase 1 (top), participants first generated five short ideas (1-3 keywords each) for eight problem statements, alternating between conditions with and without AI assistance (top left). Second, they elaborated on each idea with a one-sentence explanation, again alternating between AI-assisted and unassisted conditions (top right). Problem order and \withAI{}/\noAI{} order were counterbalanced across participants. \revision{Phase 2 (bottom) started after around one week (accessible after 6 days with a 48-hour completion window).} In Phase 2, participants saw 60 elaborations (40 self-generated and 20 distractors) together with the original problem statements. For each, they indicated whether they remembered working on this item, and if so, attributed the source of both the idea and the elaboration (self vs. AI), along with confidence ratings. Distractors included both elaborations from familiar problems (known-topic) and from unseen problems (unknown-topic). This procedure implemented a 2×2 within-subjects design (\noAI{} vs. \withAI{} during ideation and elaboration) to measure item memory and source attribution.}
    \Description{This figure presents a flow diagram of the study design. Phase 1 (top row): Participants are shown eight problems (out of a pool of ten). For each problem, they generate five short ideas (1–3 keywords each), sometimes with AI assistance and sometimes without. Icons of lightbulbs represent ideas, and robot icons indicate AI involvement. After ideation, participants elaborate on each idea by writing a one-sentence explanation, again alternating between AI-assisted and unassisted conditions. Problem order and AI/no-AI conditions are counterbalanced. Phase 2 (bottom row): After one week, participants complete a memory and source attribution task. They are shown elaborations linked to problems, including 40 items from Phase 1 and 20 distractors. Distractors come from both problems participants had seen before (known-topic) and problems they had not seen (unknown-topic). For each elaboration, participants indicate (1) whether they remember working on it, (2) whether the idea source was themselves or AI, and (3) whether the elaboration source was themselves or AI. They also provide confidence ratings. Legend: Blue icons represent participants' own elaborations from Phase 1, gray icons represent distractors from previously seen problems, and pink icons represent distractors from unseen problems. Overall, the figure illustrates the two-phase procedure designed to test item memory and source attribution under conditions with and without AI support.}
    \label{fig:study_design}
\end{figure*}

To investigate how co-writing with LLMs affects people's source memory for ideas and elaboration texts, we conducted a two-phase, within-subjects online experiment. \cref{fig:study_design} gives a visual overview.

Participants (N=\nMain{}) had to create ideas and text by themselves (\noAI) or with the help of a chatbot (\withAI) and were later asked to remember and indicate the sources.

The study was pre-registered.
Details can be accessed at: \url{https://aspredicted.org/g78p-w7db.pdf}. 
Deviations from the pre-registration can be found in \cref{tab:prereg_deviation}.

\subsection{Experiment Design} %
Our experiment consists of \textit{two phases} a week apart.
In Phase 1, participants created solution ideas and elaboration texts, either with or without AI support.
\revision{We decided to separate the writing task into two fundamental steps, ideation and elaboration, instead of letting participants solve the prescribed problems in one go. 
This structured approach made it possible to more clearly isolate the effects of using AI during idea generation versus during sentence-level writing.}
In Phase 2, they indicated the sources of idea and elaboration, given the problem and elaboration text.

Participants who completed Phase 1 received an invitation after  6 days to complete Phase 2 within 48 hours.
This delay is consistent with source memory studies in the literature (e.g., \cite{Beaufort13cryptomnesia, Bornstein1995, Brown91Cryptomnesia}) and the 48-hour time window was used to reduce drop-out.

\subsubsection{Independent Variables}
The independent variables are the presence (\withAI) or absence of AI chatbot assistance (\noAI) during ideation and during elaboration, respectively, resulting in a 2×2 within-subjects factorial design. 
Each participant experienced all four experimental conditions: 
\begin{enumerate}
    \item no AI assistance during either ideation or elaboration,
    \item AI-assisted ideation but manual elaboration,
    \item manual ideation but AI-assisted elaboration, and
    \item AI assistance during both ideation and elaboration.
\end{enumerate}
Splitting this overall content creation task into the two subtasks (ideation and elaboration) allowed us to examine how the phase in which AI support is used affects source memory.
The two subtasks also reflect common real-world use cases of LLMs: generating initial ideas (ideation) and expanding or refining ideas (elaboration).

In Phase 1, participants were asked to generate five ideas (1-3 keywords each) on how to solve diverse problems, alternating between conditions with and without AI support.
We created ten different problem statements and each participant had to work on eight of those.
The two remaining topics served as distractors in Phase 2.
To control for order effects, we counterbalanced the ideation task across participants: If one participant began the ideation phase with AI support, the next began without it.
We also counterbalanced the selection of the eight out of ten problems and their order of presentation.

Within Phase 1, after completing all ideation tasks, participants moved on to the elaboration step. 
Here, they were presented with their generated ideas in the same order in which they had created them.
For each idea, they wrote a one-sentence elaboration, again alternating between the \withAI{} and \noAI{} conditions, and counterbalanced across participants.

\revision{While we did not strictly enforce a limit of three words for ideas or one sentence for elaborations, to allow for flexibility (e.g., multi-word keywords or naturally short sentences), participants generally adhered to these guidelines:}
\revision{The median idea length was three words, and the median elaboration length was 26 words, with fewer than \pct{2.6} of all idea-elaboration pairs having identical lengths or an elaboration shorter than the idea.}

\revision{\cref{tab:example_participant_inputs} shows example inputs from the participants to illustrate the experimental process.}

\begin{table*}[]
\small
 \newcolumntype{Y}{>{\raggedright\arraybackslash}X}
\begin{tabularx}{\linewidth}{
    Y
    >{\raggedright\arraybackslash}p{2.75cm}
    >{\raggedright\arraybackslash}p{1.7cm}
    Y
    >{\raggedright\arraybackslash}p{1.7cm}
}
\toprule
\textbf{Topic/Problem} & \textbf{Idea} & \textbf{Idea source} & \textbf{Elaboration} & \textbf{Elaboration source} \\
\midrule

Think of features that could help blind or low-vision people use a city's bus/train app more easily. &
Live agent, travel guidance &
noAI &
The app could offer an option to connect with a live agent who can provide real-time travel assistance, directions, and updates to support blind or low-vision users during their commute. &
noAI \\
\midrule

Climate change is causing more frequent and severe weather events. What innovative ideas could help communities better prepare for and adapt to these changes? &
Resilient Infrastructure &
withAI &
When there's resilient infrastructures, there are top and tough buildings, roads, and utilities that can withstand extreme weather like floods, storms, and heatwaves. They reduce damage costs, keep essential services running and faster recoveries should any disaster happen &
noAI \\
\midrule

Come up with ways students can fairly distribute work for an essay in a group project. &
each writes a paragraph &
noAI &
Each student takes responsibility for writing one paragraph, ensuring equal contribution and allowing the group to combine their individual efforts into a cohesive essay. &
withAI \\
\midrule

Oceans are heavily polluted with plastic waste. What practical approaches could help reduce plastic pollution in the oceans? &
Promote eco-friendly materials &
withAI &
Promoting the use of eco-friendly materials can significantly reduce plastic waste and help protect our oceans from pollution. &
withAI \\
\bottomrule

\end{tabularx}
\caption{\revision{Example ideas and corresponding elaborations created by participants during the study.}}
\Description{This table lists four example topics used in the study, along with participants' inputs, and the sources of those ideas and elaborations.}
\label{tab:example_participant_inputs}
\end{table*}

\subsubsection{Dependent Variables}
We measured four main types of dependent variables:
\begin{enumerate}
    \item item memory performance (i.e. remembering to have worked on a problem and solution at all or not)
    \item source attribution accuracy for idea and elaboration
    \item source attribution confidence for idea and elaboration
    \item perceived task performance
\end{enumerate}

In detail, Phase 2 \revision{presented 40 items, where each item consisted of the original problem text paired with the participant's own elaboration sentence, which they created in Phase 1. 
The underlying ideas were not shown again.
In addition, we presented 20 distractor items, each of which consisted of a problem text paired with an elaboration the participant had not previously generated or seen.}
For counterbalancing, we randomised the order of problems and solutions for each participant and used this in a ``round-robin'' procedure (one idea from all problems before returning to another idea from the first problem), thus maximizing the distance between encountering solutions for the same problem. 
This was motivated by reducing context effects and shortcuts in source attribution.

\revision{Concretely,} for each elaboration shown in Phase 2, \revision{we first asked participants, ``\textit{Do you \textbf{remember} working on this solution?}''} (binary: ``Yes'' / ``No''), allowing us to assess item memory performance.
Second \revision{, we asked ``\textit{Did you come up with the \textbf{underlying idea} for the solution on your own or with AI-support?}'' and ``\textit{Did you write this \textbf{solution text} on your own or with AI-support?}''} (binary: ``On my own'' / ``With AI support'')\revision{, to measure source memory performance}.
In addition, for each attribution, they rated their confidence on a slider ranging from 0 (``Very unsure'') to 100 (``Very sure'').
Finally, at the end of Phase 2, participants estimated their own attribution accuracy for both idea and elaboration source. \revision{They could answer using a slider ranging from 0\% (``None'') to 100\% (``All''), } 
and indicated whether they believed they had more often misattributed content to themselves or to the AI.

\subsection{Participant Recruitment}
\revision{Our sample consisted of \nMain{} healthy adults.}
Out of those, 89 self-identified as female and 95 as male.
All participants were either based in the US (72 people) or the UK (112 people) and all of them listed English as their primary language, with all but 7 participants being native speakers. Their age ranged from 19 to 75 (mean=39.8, SD=12.7).
We asked participants how often they use AI tools (e.g., ChatGPT, Claude, Copilot). 
Most of them (139) reported daily use or several times a week, some (35) once a week or several times a month, and few participants (10) rarely or never use AI.
Most participants had a college degree: 
82 held a bachelor's, 28 a master's, and 7 a doctorate. 
Eleven reported an associate degree. 
37 had a high school diploma and 18 had some college education without a degree. 
One reported another type of education.

The sample was acquired through the online platform Prolific\footnote{\url{www.prolific.com}}. %
\revision{We recruited participants until we obtained \nRec{} valid and complete responses in Phase 1.
This target sample size was estimated based on an a priori power analysis, indicating that \nRec{} participants would provide \pct{80} power to detect a small correlation of $r > .20$ when calculating differences between self-rated and actual task performance, which we expected to be a small effect.} %
We excluded \nExcl{} participants from the initially recruited \nRec{} people, as they either did not complete Phase 2 of the experiment within the required 48-hour window (6 people) or our logs showed that they did not complete the tasks correctly (10 people; e.g., entering ``no idea'' or always submitting the same text).
Participants who completed both parts of the experiment received a total compensation of £23. The average completion time was 107 minutes, resulting in an effective hourly rate of approximately £12.90.
The experiment was approved by our institution's ethics committee.

\subsection{Apparatus}
Here, we describe the study materials for the tasks, followed by an overview of the experimental software.

\subsubsection{Problem Statements}
We prepared a pool of ten problem statements for the ideation task in Phase 1, spanning social, technological, and environmental issues.
We designed them to spark creative problem-solving without triggering emotional responses or appearing overly opinionated (e.g., ``List practical ways a mid-sized company could reduce employee susceptibility to phishing emails.'').
Guided by the concept of desirable difficulty~\cite{bjork2020dirabledifficulties}, we formulated the problems so that generating five ideas would be achievable for everyone within a reasonable amount of time, but challenging enough to encourage effortful cognitive processing.
A list of all ten statements can be found in \cref{sec:appendix_statements}.
For each participant, eight of those topics were used in Phase 1, and two served as distractors in Phase 2.

\subsubsection{Distractors}
To assess false memory and source misattribution, we included distractor items in Phase 2. 
These were texts that participants had not seen or generated during Phase 1. 
We implemented two types of distractors.

\paragraph{Known-topic distractors}
For each problem statement a participant had encountered in Phase 1, we added one \textit{additional elaboration sentence from the same problem} as a distractor. 
These known-topic distractors tested whether participants would falsely recognize or misattribute content that was topically familiar but otherwise novel.
To reduce the risk that \revision{these} within-topic distractors might be inadvertently similar to participants' own responses, which could lead to them ``correctly'' remembering distractors, we chose solution texts that are distinctive and always included a proper name\revision{, while still being plausible within the topic}.
\revision{Although this choice may have made these distractors comparatively easy to reject, their main purpose was to prevent participants from making decisions based solely on general topic or problem familiarity. They also provided a way to assess false recognition.}

For example, for the problem statement ``Oceans are heavily polluted with plastic waste. What practical approaches could help reduce plastic pollution in the oceans?'', the within-topic distractor was: ``Deploy AquaClean bio-magnetic kelp that's genetically modified to attract and collect microplastics for easier harvesting.''

\paragraph{Unknown-topic distractors}
We also constructed unknown-topic distractors: solution elaborations for problem statements participants had never worked on at all. 
For each of the ten possible problem statements in our pool, we created six elaboration texts: one was used as the known-topic distractor (described above), and five were rather obvious solutions that served as these unknown-topic distractors.

\subsubsection{Experimental Software}
\begin{figure}
    \centering
    \includegraphics[width=\linewidth]{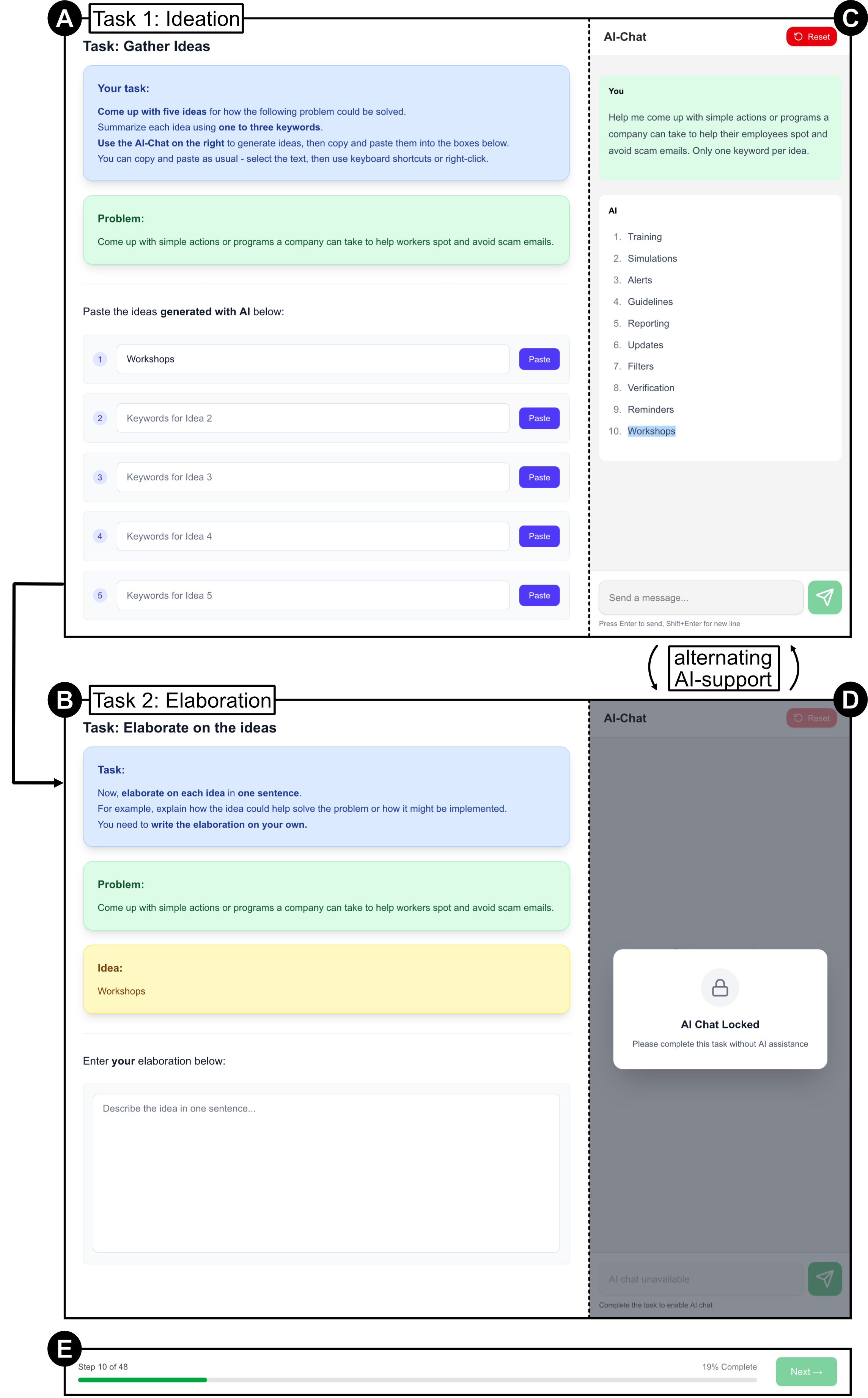}
    \caption{Creating ideas and elaborations in Phase 1: (A) In the ideation phase participants entered five ideas for each problem into the respective input fields. (B) After creating ideas for every problem, they elaborated on those ideas in one sentence each. AI-support alternated for each problem in the ideation task and each idea in the elaboration task. Therefore all participants created half of the ideas and elaborations with AI support (C) and the other half without (D).
    The progress bar (E) indicates how far along participants are in the study.}
    \Description{The figure shows a multi-panel screenshot of a web-based study interface. Panel A depicts the ideation task with input boxes for five ideas, a problem statement, and a chat panel on the right where an AI assistant provides keyword suggestions. Panel B shows the elaboration task where participants write one-sentence elaborations for each idea. Panels C and D illustrate the AI chat area, either active with AI-generated keywords (C) or locked without AI support (D). Panel E displays a progress bar at the bottom indicating 10\% completion of the study. Arrows and labels connect the panels to clarify workflow and indicate alternating AI support across tasks.}
    \label{fig:fig_frontend}
\end{figure}

We developed a custom web application using \textit{React}\footnote{\url{https://react.dev}} with \textit{Next.js}\footnote{\url{https://nextjs.org}} (\cref{fig:fig_frontend}). 
It handled communication with a PostgresDB using \textit{Prisma}\footnote{\url{https://www.prisma.io}}.
We used this database to store participant responses as well as interaction logs.
\revision{While we did not record any interaction outside of our experimental software (e.g., with external AI tools), we logged ``focus lost'' events (e.g., when particpants switched to another tab during the study) and selected keystrokes (e.g., ``paste attempts'' in the \noAI{} condition) to verify integrity.}
We used the AI-SDK\footnote{\url{https://ai-sdk.dev}} to enable communication between users and the LLM endpoint. We used the OpenAI endpoint ``gpt-4.1-mini-2025-04-14'' with the system prompt ``You are a helpful assistant.'' and provided no further context.

\begin{figure}
    \centering
    \includegraphics[width=\linewidth]{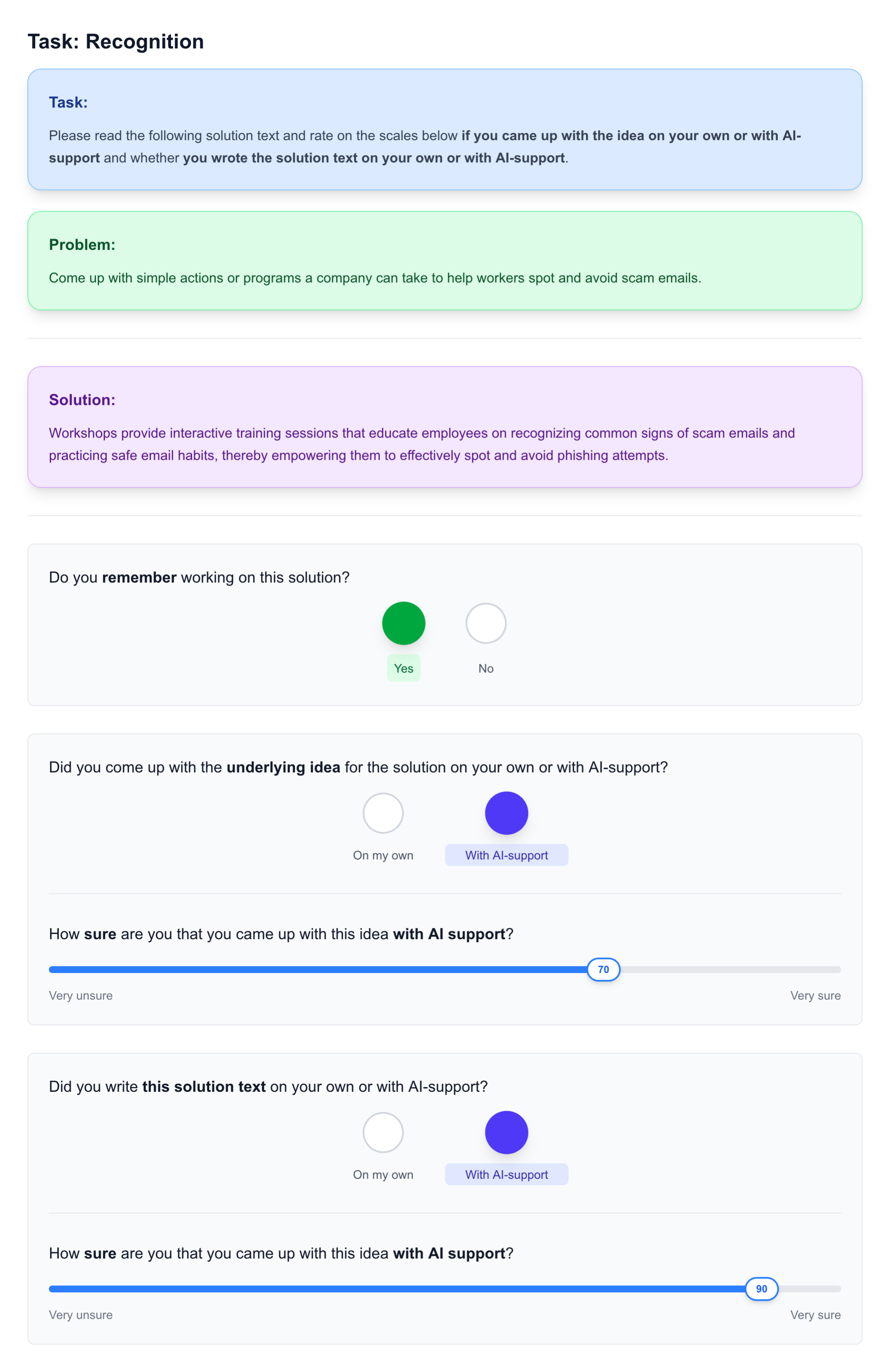}
    \caption{Memory and source attribution in Phase 2: We displayed task description, original problem statement, and solution text created in Phase 1. Participants first answered whether they remember working on a solution. If so, they indicated whether they came up with the underlying idea and the solution text on their own or with AI support. They also provided confidence ratings (0-100) for each.}
    \Description{The figure presents a web-based recognition task interface. At the top, a blue box explains the task instructions, followed by a green box showing the problem statement and a purple box displaying the solution text produced earlier. Below, participants respond to questions: a Yes/No toggle asks if they remember working on the solution, and two sets of radio buttons and sliders record whether the underlying idea and the written solution were generated independently or with AI support, along with confidence ratings from 0 to 100. The layout uses distinct colored sections and interactive elements for clarity.}
    \label{fig:fig_frontend_p2}
\end{figure}

For Phase 1, the app showed a welcome screen with study information and consent form, %
followed by the tasks (\cref{fig:fig_frontend}). In Phase 2, it then displayed the corresponding memory and attribution tasks (\cref{fig:fig_frontend_p2}).

\subsection{Procedure}
All participants completed the study online via \textit{Prolific}, using a web browser on a desktop or laptop computer. 
Mobile devices were not permitted. 
\revision{In addition, our task explicitly asked participants to use the AI chat within our experimental software (\withAI) or to solve the task on their own (\noAI).}
\revision{We found no differences between the conditions in the number of ``focus lost'' events over time or other suspicious activity. This indicates that participants generally adhered to these instructions.}

Participants first read an online information sheet and consent form, and if they agreed continued to Phase 1.
On average, participants required 74.7 minutes (SD 30.9 minutes) for Phase 1 and 32.8 minutes (SD 16.6 minutes) for Phase 2.
After completing Phase 1, they started Phase 2 after 6 days and 5 hours (SD 8 hours 46 minutes) on average.

\subsubsection{Phase 1 -- Creating Ideas and Elaborations}
The task description displayed the problem text and instructed participants to create five ideas each consisting of 1-3 keywords. 
These ideas were entered into five input fields displayed on one page (\cref{fig:fig_frontend} A).

In the \withAI{} condition, a chatbot interface was available on the right side of the screen (\cref{fig:fig_frontend} C). 
Participants could copy text from the chatbot and paste it into the input fields using the usual keyboard shortcuts \revision{(Ctrl/Cmd + C, Ctrl/Cmd + V)}, right-click or buttons located next to each field \revision{(\cref{fig:fig_frontend} A, on the right)}.
\revision{It was only possible to copy text directly from our chatbot, and manual typing was restricted to deletions (i.e. with \textit{backspace}) to ensure that participants directly used the AI output.}
In the \noAI{} condition, the chatbot was locked (\cref{fig:fig_frontend} D) and pasting \revision{any text} into the input fields was disabled.

After entering keywords into all five fields, the ``Next'' button became active, allowing participants to move on to the next ideation task. 
Each participant completed a total of eight such tasks, alternating between conditions with and without AI support, resulting in a total of 40 ideas.

Following this, participants were asked to elaborate on each of their previously submitted ideas with a one-sentence elaboration. 
Each such elaboration task displayed the task instruction, the original problem statement, and one of the participant's earlier ideas alongside a text input field (\cref{fig:fig_frontend} B).

Each elaboration page featured one idea at a time, in the order they were submitted.
Participants completed these elaborations with and without AI assistance in alternating order.

For both ideation and elaboration, the overall progress was shown at the bottom of the application (\cref{fig:fig_frontend} E).
After completing elaborations for all 40 ideas, the study concluded with a thank-you screen. 
Participants were reminded that they would receive an invitation to Phase 2 six days later and that this follow-up phase would then need to be completed within 48 hours.

\subsubsection{Phase 2 - Memory and Source Attribution}
After the six-day delay, participants who completed Phase 1 received an invitation to take part in Phase 2. 
They were presented with the study information and again provided informed consent before proceeding.

We first asked participants to guess the purpose of the study in a free text entry field.
The main part of Phase 2 then consisted of the memory and source attribution tasks (\cref{fig:fig_frontend_p2}).
Besides the overall task description, each task page displayed  the original problem statement and a solution (elaboration) text created by the participant in Phase 1.
The underlying ideas were not shown.

Participants first indicated whether they remembered working on the displayed solution, and if they did:
(1) Whether they came up with the underlying idea on their own or with AI support, (2) How confident they were in that attribution, (3) Whether they wrote the solution text on their own or with AI support, and (4) How confident they were in that second attribution.
If they did not remember working on the displayed solution at all, they skipped the attribution questions and proceeded to the next memory task.

This process was repeated for all 40 elaborated ideas from Phase 1, as well as for 8 distractor ideas related to the same problems participants had worked on (known-topic distractors), and 12 distractor ideas from two entirely new problems (unknown-topic distractors), for a total of 60.

At the end, participants filled out a final questionnaire. 
They were asked to:
(1) Estimate what percentage of ideas they believed they had correctly attributed, (2) Indicate whether they more often misattributed ideas to themselves or to the AI, (3) Estimate what percentage of solution texts they believed they had correctly attributed, and (4) Indicate whether they more often misattributed texts to themselves or to the AI.

Additionally, participants were asked to explain any strategies they used to determine the source of each idea or solution text.

Finally, participants completed a demographic questionnaire, including questions about gender, English proficiency, education level, and frequency of AI tool use.
An overview of all questions and answer options can be found in \cref{sec:appendix_questionnaire}.

Upon finishing both parts of the study, participants were shown a final thank-you screen.

\subsection{Data sharing}
We make our experimental software, study material, analysis code, and the collected data accessible through our project repository: \url{https://osf.io/dvkj9/}.
Two researchers went through all the data collected via free text input fields to ensure that all potentially privacy-sensitive information were removed before publication.

\section{Results}
We first examine people's ability and confidence in remembering content sources when working with AI.
Next, we assess participants' perceived performance.
To then disentangle underlying cognitive processes, we explore a Multinomial Processing Tree (MPT) model, allowing us to estimate separate parameters for memory, guessing, and bias in source attribution.
Finally, we report on participants' qualitative feedback to further contextualise our findings.

\subsection{Source Attribution Accuracy} %
We computed binomial generalized linear mixed models (GLLMs) in R~\cite{R2020}, using the packages \textit{lme4}~\cite{Bates2015} and \textit{lmerTest}~\cite{Kuznetsova2017}, with participants (ID) as random intercepts to test how the (ground-truth (GT)) sources of the idea and elaboration affected three types of memory: whether participants remembered (a) working on the solution at all, (b) the source of the idea, and (c) the source of the elaboration.

\begin{figure}
    \centering
    \includegraphics[width=\linewidth]{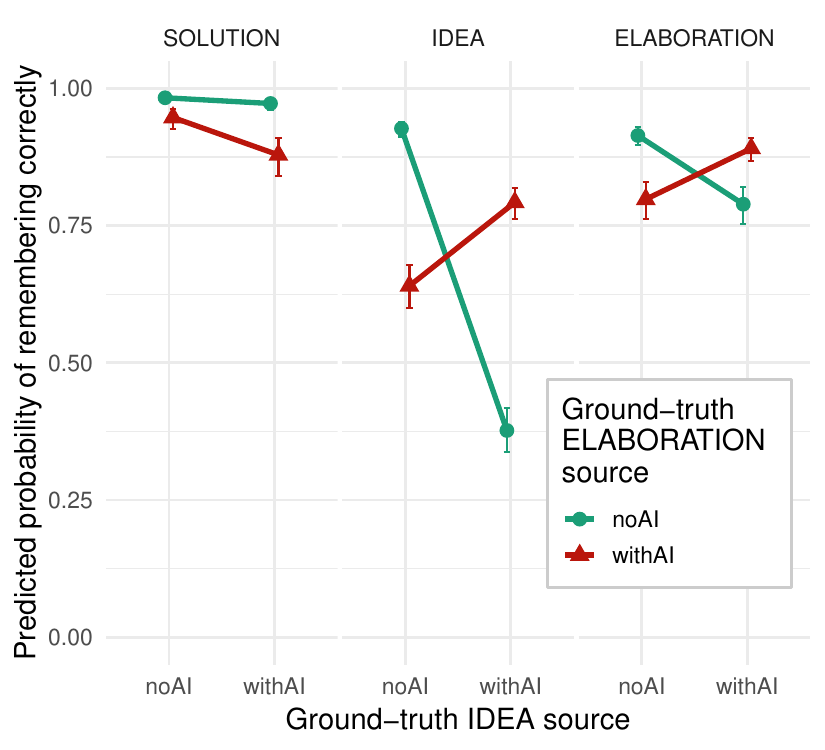}
    \caption{Model-based interaction plots of predicted source attribution accuracy for three types of memory: whether participants remembered working on the solution, remembered the source of the idea, and remembered the source of the elaboration. Probabilities are estimated from binomial GLMMs with participants as random intercepts. Error bars represent 95\% confidence intervals. The plots show that item memory was high across conditions, while idea and elaboration source memory exhibited strong interactions: Ideas created by participants without AI were better remembered when the elaboration was also created without AI. In contrast, ideas generated with AI were better remembered when AI was also used to elaborate on them.}
    \Description{This figure shows the model-based interaction plots of predicted recognition accuracy across human- and AI-sourced ideas and elaborations. The left panel shows that participants almost always remembered working on the solution, regardless of source. In contrast, the middle and right panels reveal clear interaction effects: participants more accurately remembered the source of the idea and elaboration when the elaboration's source matched the idea's source (noAI – noAI or withAI – withAI), but accuracy dropped when the sources were mixed.}
    \label{fig:combinedinteraction}
\end{figure}

\cref{fig:combinedinteraction} shows the model-based interaction plots of predicted source attribution accuracy for all three memory types across conditions (\noAI{} vs. \withAI{}).

In this section, the odds ratios (ORs) obtained from the GLMMs represent the effect sizes, quantifying how strongly the source factors influenced each type of memory.

\subsubsection{Remembering the Item} %
\revision{We first test item memory to establish whether participants recognized having worked on a solution at all.}
Overall, participants correctly remembered whether they worked on a given solution in \pct{82.4} of cases.
When they had indeed worked on the solution during Phase 1, the probability of correctly remembering this was \pct{86.9}. In contrast, the probability of correctly rejecting unseen solutions was \pct{73.4}, meaning approximately one quarter of distractors were falsely attributed as known.
Correct rejection rate varied by distractor type: \pct{60.8} for known-topic distractors and \pct{81.8} for unknown-topic distractors.

Both main effects were negative and significant, and so was the interaction: 
$\beta_{\text{idea(\withAI{})}}=-0.50$ ($p<.001$), 
$\beta_{\text{elab(\withAI{})}}=-1.19$ ($p<.001$),
$\beta_{\text{idea(\withAI{})}\times\text{elab(\withAI{})}}=-0.41$ ($p=.024$).
In odds-ratio terms, using AI for the idea reduced the odds of correctly  remembering that one had worked on the solution (OR $=0.61$), as did using AI for elaboration (OR $=0.30$); the combined effect of AI at both stages further reduced the odds (OR $=0.67$).

\cref{tab:predprob_sol} shows the model-based predicted probabilities (from \texttt{emmeans}, response scale):

\begin{table}[H]
    \centering
    \begin{tabularx}{\linewidth}{X X r r}
    \toprule
    \textbf{Idea \newline source (GT)} & \textbf{Elaboration source (GT)} & \textbf{Pr(correct) [95\% CI]} \\
    \midrule
    \noAI{} & \noAI{} & \textbf{97.6}\% [97.6, 98.8]\\
    \withAI{}    & \noAI{} & 97.3\% [96.1, 98.1]\\
    \noAI{} & \withAI{}    & 94.7\% [92.7, 96.2]\\
    \withAI{}    & \withAI{}    & 87.9\% [84.0, 90.9]\\
    \bottomrule
    \end{tabularx}
    \caption{Model-based predicted probabilities for correctly remembering having worked on a solution at all.}
    \Description{The table lists model-based predicted probabilities of participants correctly remembering that they worked on a solution. Four rows represent combinations of idea source and elaboration source, either without AI (noAI) or with AI (withAI).}
    \label{tab:predprob_sol}
\end{table}

Pairwise contrasts (Holm-adjusted) showed that the no-AI workflow (\noAI{}/\noAI{}) yielded higher accuracy than any condition with AI: 
vs.\ \withAI{}/\noAI{}, OR $=2.02$; 
vs.\ \noAI{}/\withAI{}, OR $=3.29$; 
vs.\ \withAI{}/\withAI{}, OR $=8.15$, all $p<.001$. 
Among those, \withAI{}/\noAI{} exceeded \noAI{}/\withAI{} (OR $=1.99$) and \withAI{}/\withAI{} (OR $=4.01$); 
\noAI{}/\withAI{} also exceeded \withAI{}/\withAI{} (OR $=2.45$); again all $p<.001$. 
These results indicate that both AI idea generation and AI elaboration independently impair memory, and their joint use compounds this effect, with the strongest decrement observed when AI contributed at the elaboration stage.

\subsubsection{Remembering the Source of Ideas} %
\revision{Next, to examine idea-source memory, we analyse how well participants differentiate between self- and AI-generated ideas.}
Overall, participants correctly indicated the source of the ideas used in their solutions created during Phase~1 in \pct{65.8} of cases.

Both main effects were negative and significant, qualified by a strong positive interaction: $\beta_{\text{idea(\withAI{})}}=-3.05$, 
$\beta_{\text{elab(\withAI{})}}=-1.97$, 
$\beta_{\text{idea(\withAI{})}\times\text{elab(\withAI{})}}=3.81$; all $p<.001$. 
That is, using AI for the idea alone sharply reduced correct attribution (OR $=0.05$), as did using AI only for elaboration (OR $=0.14$); when both idea \emph{and} elaboration came from AI, the interaction increased the odds ($\times 45.3$), partially offsetting the main-effect penalties.
Model-based predicted probabilities (\cref{tab:predprob_idea}) make the pattern clear:

\begin{table}[H]
    \centering
    \begin{tabularx}{\linewidth}{X X r r}
    \toprule
    \textbf{Idea \newline source (GT)} & \textbf{Elaboration source (GT)} & \textbf{Pr(correct) [95\% CI]} \\
    \midrule
    \noAI{} & \noAI{} & \textbf{92.4}\% [91.5, 93.2]\\
    \withAI{}    & \noAI{} & 37.7\% [33.6, 41.9]\\
    \noAI{} & \withAI{}    & 64.0\% [60.2, 67.6]\\
    \withAI{}    & \withAI{}    & 79.3\% [75.9, 82.4]\\
    \bottomrule
    \end{tabularx}
    \caption{Model-based predicted probabilities for correctly remembering the idea source.}
    \label{tab:predprob_idea}
    \Description{The table presents predicted probabilities of participants correctly recalling whether the idea was generated with or without AI support. Four rows represent combinations of idea source and elaboration source, either without AI (noAI) or with AI (withAI).}
\end{table}

Participants were most accurate without AI (\noAI{}/\noAI{}), and least accurate when the idea came from AI but participants elaborated themselves (\withAI{}/\noAI{}). 
Notably, keeping the workflow consistent (\withAI{}/\withAI{}) substantially improved accuracy relative to the two mixed workflows (79.3\% vs.\ 37.7\% and 64.0\%), indicating that mismatched sources between ideation and elaboration particularly impair memory for where the idea originated.

\subsubsection{Remembering the Source of Elaborations}
\revision{To examine elaboration source memory, we now analyse how well participants remember whether they wrote sentences by themselves or together with AI.}
Participants correctly indicated the source of the elaborations they had created during Phase 1 in \pct{72.8} of cases.

Both main effects were negative and significant, qualified by a strong positive interaction: $\beta_{\text{idea(\withAI{})}}=-1.05$, 
$\beta_{\text{elab(\withAI{})}}=-0.99$, $\beta_{\text{idea(\withAI{})}\times\text{elab(\withAI{})}}=1.77$; all $p<.001$. 
As such, using AI for the idea alone reduced correct elaboration-source attribution (OR $=0.35$), as did using AI only for elaboration (OR $=0.37$); when AI was used for both idea \emph{and} elaboration, the interaction increased the odds ($\times 5.90$), partially compensating for the main effects.

Model-based predicted probabilities (\cref{tab:predprob_elab}) were:

\begin{table}[H]
    \centering
    \begin{tabularx}{\linewidth}{X X r r}
    \toprule
    \textbf{Idea \newline source (GT)} & \textbf{Elaboration source (GT)} & \textbf{Pr(correct)} [95\% CI]\\
    \midrule
    \noAI{} & \noAI{} & \textbf{91.5}\% [90.1, 92.7]\\
    \withAI{}    & \noAI{} & 79.0\% [76.3, 81.5]\\
    \noAI{} & \withAI{}    & 80.0\% [77.3, 82.5]\\
    \withAI{}    & \withAI{}    & 89.2\% [87.2, 91.0]\\
    \bottomrule
    \end{tabularx}
    \caption{Model-based predicted probabilities for correctly remembering the elaboration source.}
    \Description{The table presents predicted probabilities of participants correctly recalling whether the elaboration was generated with or without AI support. Four rows represent combinations of idea source and elaboration source, either without AI (noAI) or with AI (withAI).}
    \label{tab:predprob_elab}
\end{table}

Participants were most accurate when no AI was involved at all (\noAI{}/\noAI{}) or when both idea and elaboration came from AI (\withAI{}/\withAI{}), with accuracy dropping in mixed workflows. 
Accuracy was lowest when AI provided the idea but participants elaborated themselves (\withAI{}/\noAI{}) or when participants provided the idea but AI elaborated (\noAI{}/\withAI{}).

\subsubsection{Summary}
Across all three memory measures, AI involvement reduced accuracy, though the magnitude and nature of this effect differed by memory type:

For \emph{remembering having worked on the solution at all} (item memory), performance was generally high yet AI use lowered the odds of correct recognition by about 40\% for ideation and by about 70\% for elaboration.

For \emph{remembering the idea source}, accuracy was substantially lower.
Participants were most accurate when no AI was involved.
Using AI consistently (\withAI{}/\withAI{}) was less accurate but higher than mixed workflows:
Using AI for only the idea or only the elaboration reduced the odds, with \withAI{}/\noAI{} being the least accurate.  

For \emph{remembering the elaboration source}, the ranking of conditions was identical to idea source memory, but overall accuracy was higher.

\subsection{Source Attribution Confidence} %
We used LMMs with participant ID as a random intercept to examine how the ground-truth (GT) sources of ideas and elaborations shaped participants' \emph{confidence} in their attributions.
Analyses focused separately on (a) confidence in remembering the idea source and (b) confidence in remembering the elaboration source.

Afterwards, we investigate the effect of correctness on confidence, by looking at the pairwise contrasts of correct vs. incorrect attributions, to examine participants' \emph{metacognitive sensitivity}.

Lastly, we use Pearson point-biserial correlation to examine how participants' confidence correlates with their source attribution accuracy.

As confidence was measured on a continuous 0–100 scale, we report model-based estimated marginal means (EMMs) and regression coefficients ($\beta$) rather than odds ratios. 
These values serve as effect sizes in this section, quantifying how strongly the source factors influenced participants' confidence in their attributions.

\begin{figure}
    \centering
    \includegraphics[width=\linewidth]{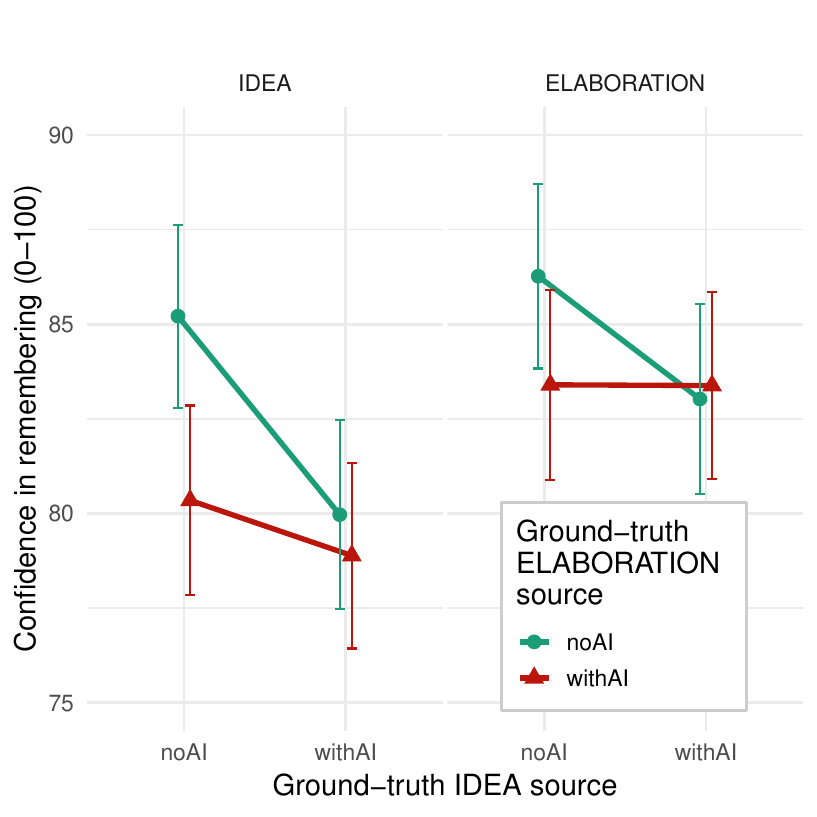}
    \caption{Confidence by idea and elaboration ground truth sources. Points show estimated marginal means; error bars show adjusted 95\% CIs.}
    \Description{The figure is a two-panel line plot displaying participants’ confidence (0–100) in remembering idea and elaboration sources. The left panel shows confidence for the ground-truth idea source, and the right panel for the ground-truth elaboration source. Each panel plots two lines: a green line with circular markers for elaborations created without AI (noAI) and a red line with triangular markers for elaborations created with AI (withAI). Confidence is higher when both idea and elaboration are noAI, and lower when the idea is withAI. Error bars represent adjusted 95\% confidence intervals.}
    \label{fig:fig_conf_GT}
\end{figure}

As an overview, \cref{fig:fig_conf_GT} shows interaction plots of model-based mean confidence in idea and elaboration attribution.

\subsubsection{Confidence in Remembering the Idea Source}
\revision{After evaluating accuracy, we next assess participants' confidence in idea-source attributions to see how well participants are able to judge their own ability to remember whether an idea originated from AI vs. themselves.}
For confidence in remembering the idea source, the grand mean across conditions was 81.10 (95\% CI [79.29, 82.91]). 
Confidence was highest when both idea and elaboration originated without AI support (EMM~=~85.2). 
Confidence dropped when AI was involved in either stage (EMMs = 80.0 and 80.3) and was lowest when AI was used in both stages (78.9). 
A significant Idea~$\times$~Elaboration interaction ($\beta=0.95$, $p<.001$) indicated that the reduction with AI involvement was not purely additive.
\cref{fig:fig_conf_GT} (left) shows the interaction.

The estimated marginal means (\cref{tab:emm_idea}, from \texttt{emmeans}, 0--100 scale) were:

\begin{table}[H]
    \centering
    \begin{tabularx}{\linewidth}{X X r r}
    \toprule
    \textbf{Idea \newline source (GT)} & \textbf{Elaboration \newline source (GT)} & \textbf{Confidence [95\% CI]} \\
    \midrule
    \noAI{} & \noAI{} & \textbf{85.2} [82.8, 87.6] \\
    \withAI{} & \noAI{} & 80.0 [77.5, 82.5] \\
    \noAI{} & \withAI{} & 80.3 [77.8, 82.9] \\
    \withAI{} & \withAI{} & 78.9 [76.4, 81.3] \\
    \bottomrule
    \end{tabularx}
    \caption{Model-based estimated marginal means for confidence in remembering the \textit{idea} source.}
    \Description{The table reports model-based estimated marginal means for participants' confidence in remembering the idea source, with values on a 0–100 scale. Four rows represent combinations of idea source and elaboration source, either without AI (noAI) or with AI (withAI).}
    \label{tab:emm_idea}
\end{table}

Pairwise contrasts (Holm-adjusted) showed that the no-AI workflow (\noAI{}/\noAI{}) yielded significantly higher confidence than any AI-involved condition ($p<.001$). 
Among those, differences were generally small, with only a modest advantage of \noAI{}/\withAI{} over \withAI{}/\withAI{} ($p=.024$).

\subsubsection{Confidence in Remembering the Elaboration Source}
\revision{We now analyse confidence in elaboration-source judgments to determine whether metacognitive insight differs between ideation and elaboration.}
For elaboration source confidence, the grand mean across conditions was 84.0 (95\% CI [82.2, 85.9]). 
Confidence was highest when both idea and elaboration were created without AI (EMM~=~86.3). 
It dropped when AI was involved at either stage: ideas ($\beta=-0.82$, $p<.001$) and elaborations ($\beta=-0.63$, $p<.001$). 
A significant interaction ($\beta=0.81$, $p<.001$) indicated that when both stages were AI, confidence was higher than expected from adding the two main effects.
\cref{fig:fig_conf_GT} (right) shows the interaction.

The model-based estimated marginal means (\cref{tab:emm_elab}) were:

\begin{table}[H]
    \centering
    \begin{tabularx}{\linewidth}{X X r r}
    \toprule
    \textbf{Idea \newline source (GT)} & \textbf{Elaboration \newline source (GT)} & \textbf{Confidence [95\% CI]} \\
    \midrule
    \noAI{} & \noAI{} & \textbf{86.3} [83.8, 88.7] \\
    \withAI{} & \noAI{} & 83.0 [80.5, 85.5] \\
    \noAI{} & \withAI{} & 83.4 [80.9, 85.9] \\
    \withAI{} & \withAI{} & 83.4 [80.9, 85.8] \\
    \bottomrule
    \end{tabularx}
    \caption{Model-based estimated marginal means for confidence in remembering the \textit{elaboration} source.}
    \Description{The table reports model-based estimated marginal means for participants' confidence in remembering the elaboration source, with values on a 0–100 scale. Four rows represent combinations of idea source and elaboration source, either without AI (noAI) or with AI (withAI).}
    \label{tab:emm_elab}
\end{table}

Pairwise contrasts (Holm-adjusted) confirmed that the all-human workflow (\noAI{}/\noAI{}) elicited significantly greater confidence than any AI-involved workflow ($p<.001$). 
By contrast, we found no significant differences between the AI-involved workflows.

\subsubsection{Metacognitive Sensitivity}
Metacognitive sensitivity describes the ability to accurately evaluate one's own decisions, thoughts, or perceptions by distinguishing between correct and incorrect responses.
\revision{Here, we compare confidence for correct versus incorrect responses, allowing us to assess how well participants can monitor their memory performance in human-AI workflows.}
We operationalized this as the difference in confidence between (ground truth) \emph{correct} and \emph{incorrect} attributions (Correct $-$ Incorrect), estimated with \texttt{emmeans} from the integrated models that included ground truth and correctness factors. 
Positive values indicate higher confidence when the attribution was correct. 

\paragraph{Idea source confidence}
Metacognitive sensitivity varied strongly by GT combination, as can be seen in \cref{tab:metacog_deltas_ideas}.

\begin{table}[H]
    \centering
    \begin{tabularx}{\linewidth}{X X r r}
    \toprule
    \textbf{Idea \newline source (GT)} & \textbf{Elaboration \newline source (GT)} & \textbf{Idea $\Delta$ [95\% CI]} & \textbf{$p$} \\
    \midrule
    \noAI{} & \noAI{} & \textbf{10.57} [7.79, 13.34] & <.0001 \\
    \withAI{} & \withAI{} & 5.04 [2.60, 7.48] & .0001 \\
    \noAI{} & \withAI{} & 1.65 [-1.39, 4.70] & .288 \\
    \withAI{} & \noAI{} & 0.83 [-1.43, 3.09] & .472 \\
    \bottomrule
    \end{tabularx}
    \caption{Metacognitive sensitivity ($\Delta$ = Correct $-$ Incorrect) by ground truth combination. Values are in confidence-scale points; Idea $\Delta$s are significant for \noAI{}/\noAI{} and \withAI{}/\withAI{}, and not significant for mixed workflows.}
    \Description{The table reports model-based estimates of metacognitive sensitivity (delta = Correct − Incorrect) in confidence-scale points. Four rows represent combinations of idea source and elaboration source, either without AI (noAI) or with AI (withAI).}
    \label{tab:metacog_deltas_ideas}
\end{table}

\begin{figure}[t]
  \centering
  \includegraphics[width=\linewidth]{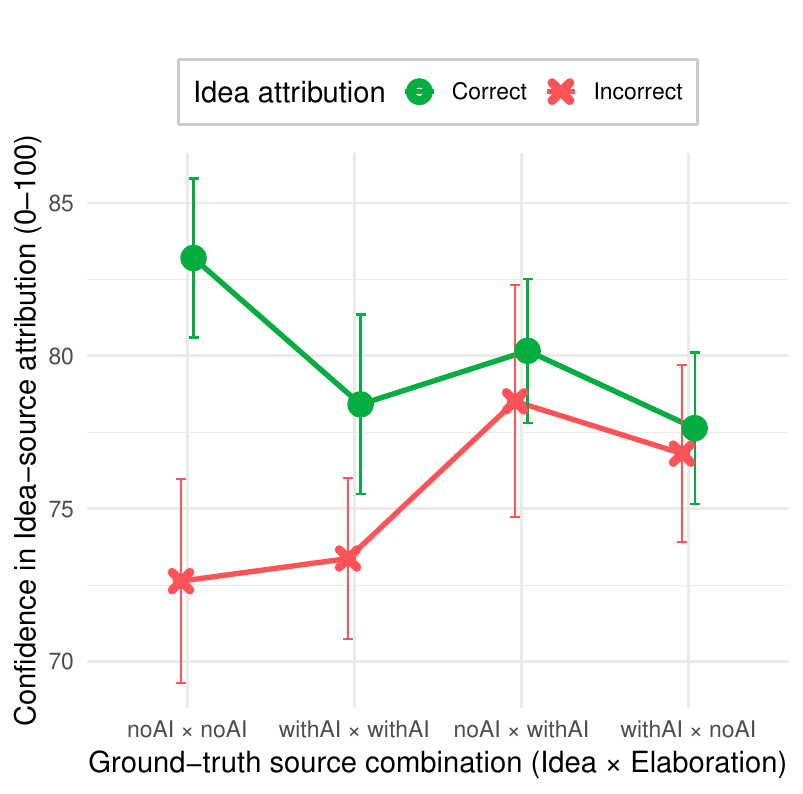}
  \caption{Metacognitive sensitivity plots for \textit{idea}: confidence for \emph{Correct} vs.\ \emph{Incorrect} attributions within each ground truth combination. 
  Points are estimated marginal means with the same (Holm-adjusted) 95\% CIs as used elsewhere. 
  \textit{Idea}-source confidence shows a large correctness gap in \noAI{}/\noAI{}, a moderate gap in  \withAI{}/\withAI{}, and minimal gaps in mixed workflows.}
  \Description{The figure displays a line graph of metacognitive sensitivity for idea confidence, comparing correct versus incorrect source attributions across four ground-truth idea–elaboration combinations. The x-axis lists the combinations (noAI × noAI, withAI × noAI, noAI × withAI, withAI × withAI), and the y-axis shows confidence in idea-source attribution (0–100). Green circles represent correct attributions and red crosses represent incorrect ones, with vertical error bars showing confidence intervals. The largest confidence gap between correct and incorrect occurs for noAI × noAI, a moderate gap appears for withAI × withAI, and minimal gaps are visible in the mixed-source conditions.}
  \label{fig:fig_cc_idea}
\end{figure}

Pairwise comparisons (Holm-adjusted) showed that \noAI{}/\noAI{} exhibited a larger sensitivity than \withAI{}/\noAI{} (difference in $\Delta=9.74$, $p<.0001$), \noAI{}/\withAI{} ($8.92$, $p=.0001$), and \withAI{}/\withAI{} ($5.53$, $p=.0136$); \withAI{}/\withAI{} exceeded \withAI{}/\noAI{} ($4.21$, $p=.0366$); other differences were not significant. 
Thus, participants' confidence tracked correctness most strongly without AI, moderately when using AI for both idea and elaboration, and weakly in mixed workflows.
\cref{fig:fig_cc_idea} shows the confidence for \emph{Correct} vs.\ \emph{Incorrect} attributions within each ground truth combination.

\paragraph{Elaboration source confidence}

Here, metacognitive sensitivity was positive in all cases, as can be seen in \cref{tab:metacog_deltas_elabs}.

\begin{table}[H]
    \centering
    \begin{tabularx}{\linewidth}{X X r r}
    \toprule
    \textbf{Idea \newline source (GT)} & \textbf{Elaboration \newline source (GT)} & \textbf{Elaboration $\Delta$ [95\% CI]} & \textbf{$p$} \\
    \midrule
    \noAI{} & \noAI{} & 5.54 [2.99, 8.09] & <.0001 \\
    \withAI{} & \withAI{} & \textbf{10.95} [8.70, 13.20] & <.0001 \\
    \noAI{} & \withAI{} & 6.47 [3.66, 9.28] & <.0001 \\
    \withAI{} & \noAI{} & 7.79 [5.71, 9.88] & <.0001 \\
    \bottomrule
    \end{tabularx}
    \caption{Metacognitive sensitivity ($\Delta$ = Correct $-$ Incorrect) by ground truth combination. Values are in confidence-scale points; All elaboration $\Delta$s are significant.}
    \Description{The table reports model-based estimates of metacognitive sensitivity (delta = Correct − Incorrect) in confidence-scale points. Four rows represent combinations of idea source and elaboration source, either without AI (noAI) or with AI (withAI).}
    \label{tab:metacog_deltas_elabs}
\end{table}

\begin{figure}[t]
  \centering
  \includegraphics[width=\linewidth]{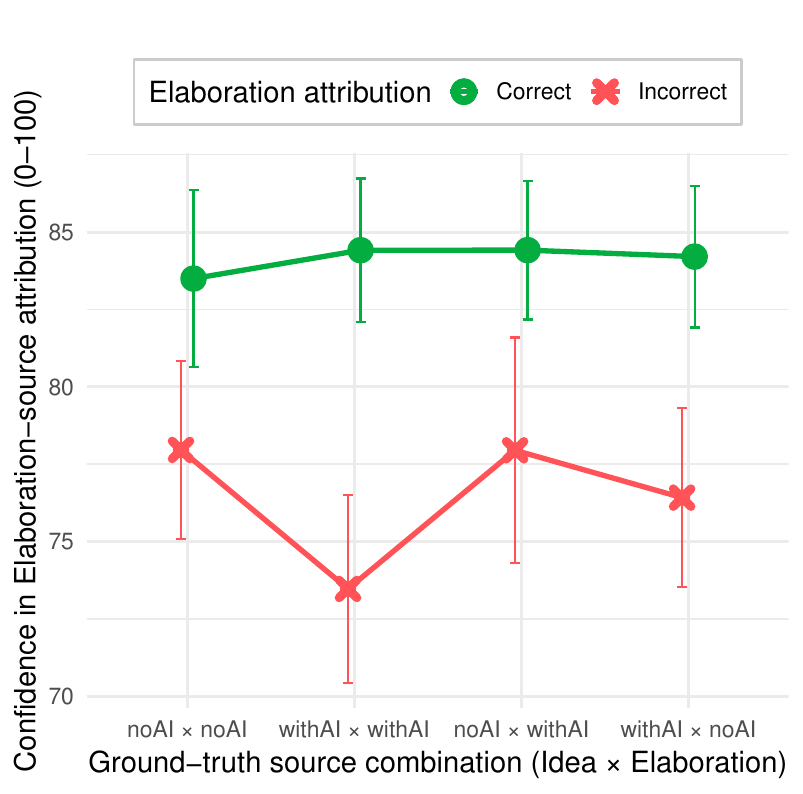}
  \caption{Metacognitive sensitivity plots \textit{for elaboration}: confidence for \emph{Correct} vs.\ \emph{Incorrect} attributions within each ground truth combination. 
  Points are estimated marginal means with the same (Holm-adjusted) 95\% CIs as used elsewhere. 
  \textit{Elaboration}-source confidence shows correctness gaps in all cells, largest in  \withAI{}/\withAI{}.
  }
  \Description{The figure is a line plot of metacognitive sensitivity for elaboration confidence, comparing correct versus incorrect source attributions across four ground-truth idea–elaboration combinations. The x-axis shows combinations (noAI × noAI, withAI × noAI, noAI × withAI, withAI × withAI), and the y-axis represents confidence in elaboration-source attribution from 0 to 100. Green circles with error bars indicate correct attributions and red crosses with error bars show incorrect attributions, with confidence intervals. All conditions display a positive gap where correct attributions receive higher confidence, with the largest difference in the withAI × withAI condition.}
  \label{fig:fig_cc_elab}
\end{figure}

Contrasts (Holm) indicated a larger sensitivity in  \withAI{}/\withAI{} than \noAI{}/\noAI{} (difference in $\Delta=5.41$, $p=.0112$); the remaining differences among AI-involved workflows were not significant (all $p>.17$). Thus, participants' confidence in elaboration attribution tracked correctness in all workflows, with the highest sensitivity when AI was used for both idea and elaboration.
\cref{fig:fig_cc_elab} shows the confidence for \emph{Correct} vs.\ \emph{Incorrect} attributions within each ground truth combination.

\subsubsection{Confidence-Accuracy Correlation}
\revision{Finally, we correlate confidence with accuracy to evaluate overall calibration and determine whether confidence reliably predicts correctness across workflows.}
We quantified \revision{this} \emph{metacognitive calibration} by correlating trial-level confidence with attribution accuracy (Correct~$=1$, Incorrect~$=0$) using Pearson point-biserial correlations. 
We report overall effects and effects \emph{within} each ground truth combination of idea and elaboration sources (all $p$-values after Holm correction).

\paragraph{Overall calibration}
Confidence reliably tracked correctness for both outcomes, with stronger coupling for elaborations than ideas ($r_{\text{idea}}=.115$; $r_{\text{elab}}=.218$; both $p<.001$).
Thus, participants' confidence was a small but dependable indicator of being correct, especially for elaboration-source judgments.

\paragraph{By ground-truth combination}
For \textit{idea} judgments, calibration depended on the workflow: 
Confidence tracked accuracy in consistent workflows (\noAI{}/\noAI{} $r=.228$; \withAI{}/\withAI{} $r=.234$; both $p<.001)$, but not in mixed workflows: 
\noAI{}/\withAI{} was essentially zero ($r=-.001$, $p=.976$), and \withAI{}/\noAI{} showed a small negative association ($r=-.094$, $p=.0011$). %

For \textit{elaboration} judgments, confidence-accuracy coupling was positive and significant in all cases. 
It was strongest for consistent workflows, particularly \withAI{}/\withAI{} ($r=.277$), and more modest in mixed workflows (\noAI{}/\withAI{}: $r=.169$ ; \withAI{}/\noAI{}: $r=.192$).

\cref{tab:corr_bygt} gives an overview.

\begin{table}[H]
\centering
\begin{tabularx}{\linewidth}{p{1.5cm} X r r}
\toprule
\textbf{Outcome} & \textbf{(GT) Idea $\times$ \newline Elaboration} & $\mathbf{r}$ [\textbf{95\% CI}] & \textbf{$p$}\\
\midrule
Idea & \noAI{}/\noAI{} & .228 [.186, .268] & $<.001$ \\
Idea & \withAI{}/\withAI{} & .234 [.189, .278] & $<.001$ \\
Idea & \noAI{}/\withAI{} & $\approx 0$ [$-.056$, .054] & .976 \\
Idea & \withAI{}/\noAI{} & $-.094$ [$-.147$, $-.041$] & .0011 \\
\midrule
Elaboration & \noAI{}/\noAI{} & .215 [.173, .256] & $<.001$ \\
Elaboration & \withAI{}/\withAI{} & \textbf{.277} [.232, .320] & $<.001$ \\
Elaboration & \noAI{}/\withAI{} & .169 [.115, .222] & $<.001$ \\
Elaboration & \withAI{}/\noAI{} & .192 [.140, .243] & $<.001$ \\
\bottomrule
\end{tabularx}
\caption{Point-biserial correlations between confidence and accuracy in each GT combination. Holm-adjusted $p$-values control the familywise error rate within each outcome.}
\Description{The table reports point-biserial correlations between confidence and accuracy within each ground-truth idea–elaboration combination. Four rows per outcome (Idea and Elaboration) represent combinations of idea source and elaboration source, either without AI (noAI) or with AI (withAI), for a total of eight rows.}
\label{tab:corr_bygt}
\end{table}

\subsubsection{Summary}
\emph{AI involvement reduced confidence} in source attributions, with the highest confidence in all-human workflows.

For \emph{idea-source confidence}, consistency mattered: 
human-only workflows were highest, fully AI-supported workflows lowest, and mixed workflows in between.
For \emph{elaboration-source confidence}, all AI-involved workflows showed similarly reduced confidence.

\emph{Correctness boosted confidence} (metacognitive sensitivity), but patterns differed: 
strong for ideas created without AI, moderate with full AI-support, and negligible in mixed workflows; 
for elaborations, sensitivity was positive in all workflows, largest with full AI.

Overall, \emph{confidence was a reliable but modest predictor of accuracy}, with stronger calibration for elaborations than ideas.

\subsection{Objective vs. Perceived Performance} %
\revision{To examine how well participants evaluated their own memory performance, we compared their actual source-attribution accuracy with their self-assessed performance. This reveals whether participants' self-evaluations aligned with their actual performance or showed systematic misjudgment.}
For this analysis, we first computed per-participant accuracy rates for both idea source and elaboration source attribution as follows:
For items seen in Phase 1, a response was correct if the item was recognised as seen and the reported source matched the ground truth. 
For distractor items, a response was correct if participants correctly detected that they had not worked on this in Phase 1.

We then compared these objective performance rates to participants' self-assessments, collected after the recognition tasks.
We used Spearman rank correlations to assess the correlation between actual and perceived performance.
Because the self-assessments were expressed as percentages (0-100\%), we rescaled them to the 0-60 range used for the objective measure (i.e., number of correct attributions out of the max of 60) for comparability.

\paragraph{Correlations between objective and perceived performance}
For ideas, the association between source attribution accuracy and self-assessment was weak and not significant, $\rho = .13$ ($p = .077$). 
By contrast, it was significant for elaborations, $\rho = .25$ ($p < .001$).

\paragraph{Calibration ratios}
Calibration ratios (self-rated / objective performance) exceeded 1 for both tasks, indicating general overconfidence: 
idea, $M = 1.12$, $t(183) = 3.81$ ($p < .001$); 
elaboration, $M = 1.06$, $t(183) = 2.03$ ($p = .043$). 
Ratios were significantly higher for idea than elaboration, $t(183) = 3.22$ ($p = .0015$).

\begin{figure}[t]
\centering
\includegraphics[width=\linewidth]{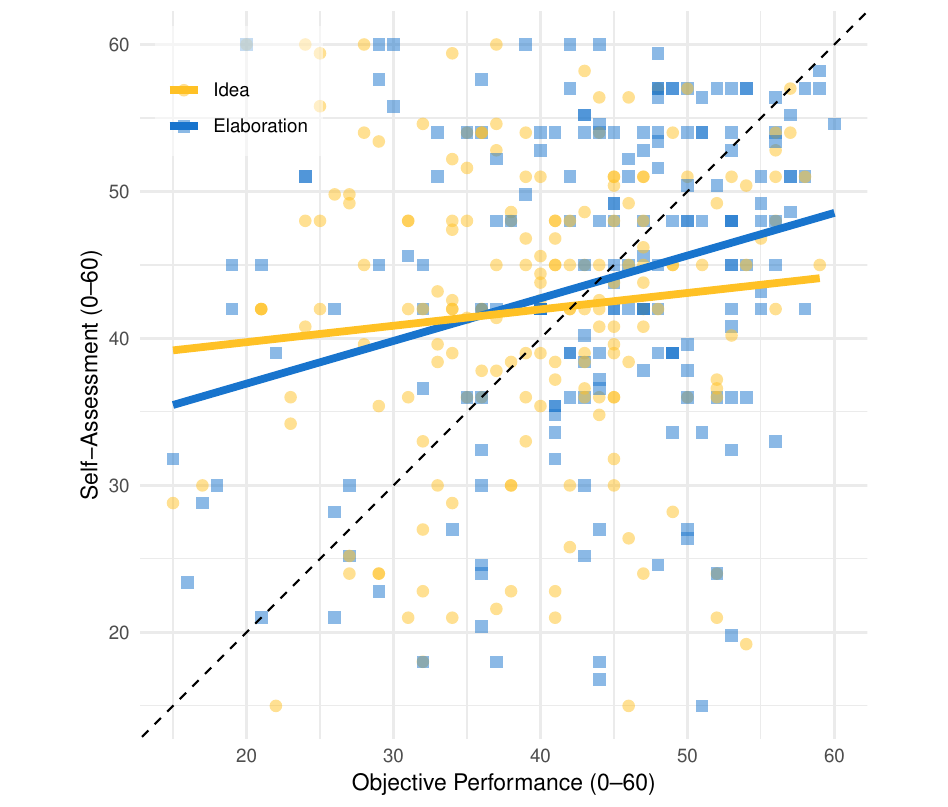}
\caption{Relationship between objective accuracy and self-assessed performance for idea- and elaboration-source attribution. The dashed line represents perfect calibration.}
\Description{The figure is a scatter plot comparing objective performance (0–60) on the x-axis with self-assessed performance (0–60) on the y-axis for idea- and elaboration-source attribution. Yellow circles represent idea data and blue squares represent elaboration data. Solid yellow and blue regression lines show the relationship for idea and elaboration, respectively. A black dashed diagonal line indicates perfect calibration where self-assessment equals objective performance. Both regression lines have positive slopes, indicating higher self-assessments with higher objective scores, though many points lie above the dashed line, suggesting overestimation.}
\label{fig:perf_vs_self}
\end{figure}

\cref{fig:perf_vs_self} plots objective accuracy against participants' self-reports for each measure.

Together, these findings suggest that participants' metacognitive monitoring of source attribution performance was more accurate for elaborations than for ideas. 
Elaboration self-assessments almost tracked actual performance and showed minimal bias, whereas idea self-assessments were weakly related to performance and systematically overestimated accuracy. 
Across both tasks, participants showed a tendency toward overestimating their performance, more strongly for ideas than elaborations.

\subsection{Multinomial Processing Tree (MPT) Model}
\label{sec:mpt} %
Recognition experiments with categorical responses are better analysed
with MPT models \cite{broder2007measuring}. For a tutorial, see \citet{schmidt2023develop}.
An MPT specifies a small set of latent cognitive states (e.g.\ ``detect old'', ``remember source'', ``guess'') and the probabilistic paths that link those states to the observed response frequencies. \revision{We use ``old'' and ``new'' in a standard recognition-memory sense: \textit{old} items are the solution texts that participants had previously created in Phase 1, whereas \textit{new} items refer to the distractors (known-topic or unknown-topic), only shown in Phase 2.}
MPTs thus serve as computational models of recognition memory. %
Compared with raw proportion tests or LMMs, MPTs disentangle cognitive memory processes from pure response and yield parameter estimates that are directly interpretable in psychological terms. \revision{In line with standard applications of the MPT, our model assumes that, within each component (idea vs elaboration), each participant is characterised by a set of latent probabilities ($d, s, \beta, f$) that apply across the set of items in that component. Thus, the MPT does not attempt to model a single solution separately but rather captures each participant's average detection, source-memory, and guessing tendencies over the specific ideas and elaborations included in the task.}

\begin{figure*}
    \centering
    \includegraphics[width=\linewidth]{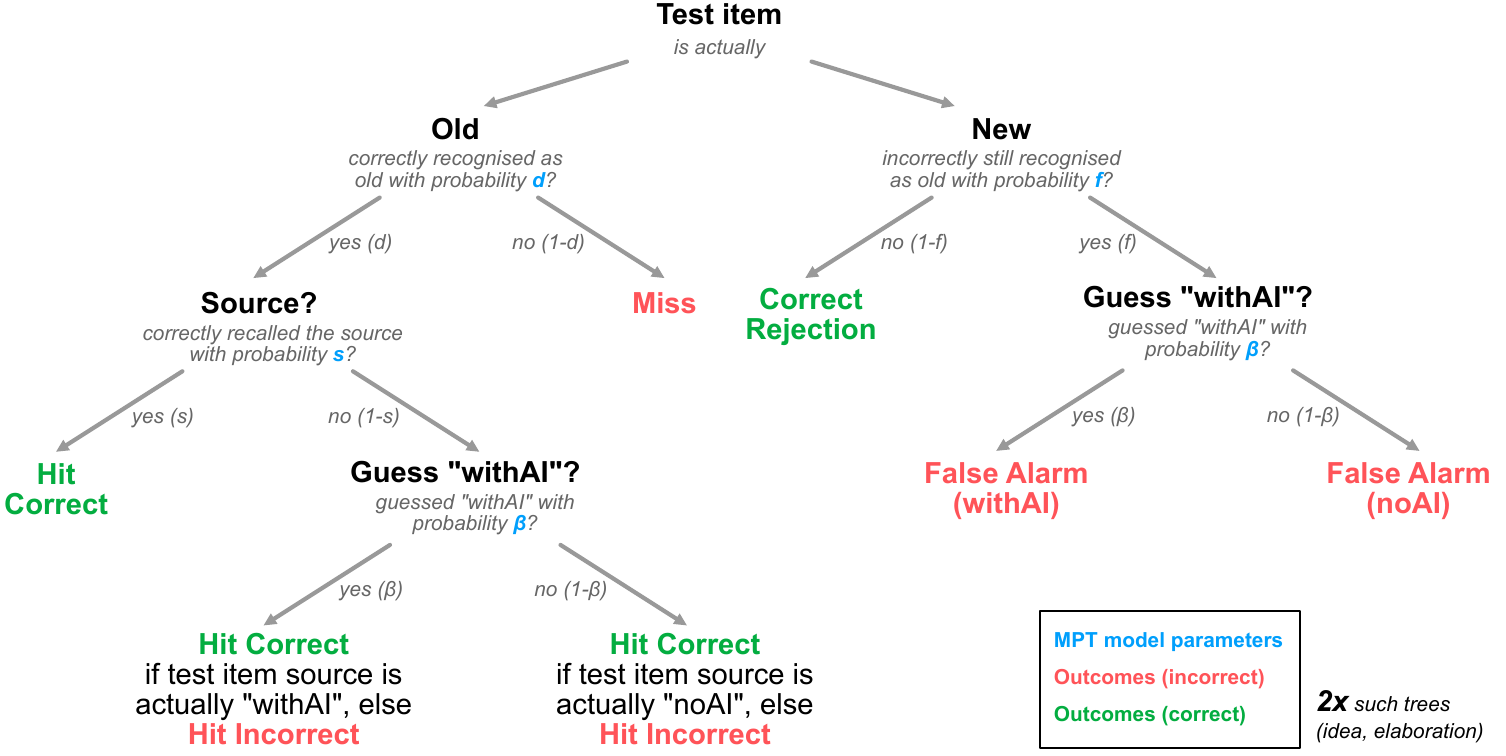}

    \caption{Overview of the MPT model: After an item appears at test, $d$ is the probability that the participant recognises it as \textit{old}. A higher $d$ therefore reflects greater mnemonic sensitivity, independent of any knowledge about the item's source. When recognition of an old item fails ($1-d$), the model routes the trial directly to the \texttt{Miss} category. Note that the earlier GLMM outcome ``remembered working on it'' is not identical to the MPT detection parameter $d$: Whereas $d$ represents a latent probability of correctly detecting studied items, the GLMM measure pools across hits, misses, and distractor rejections, so the two indices are not directly comparable.
    When the source is forgotten, the model assumes the participant
    guesses. Parameter $\beta$ is the probability that this guess is ``\withAI{}''; $1-\beta$ is therefore the probability of guessing ``\noAI{}''. Values of $\beta$ above $0.5$ indicate a systematic bias toward attributing the item to the AI system, whereas $\beta=0.5$ corresponds to an unbiased coin flip.
    On distractor trials there is, by definition, no true episodic trace. Parameter $f$ is the probability of producing an \textit{old} response nevertheless. The ensuing attribution is, again, governed by $\beta$, yielding the \texttt{False Alarm (\withAI)} or \texttt{False Alarm (\noAI)} categories. The complement $1-f$ leads to the category \texttt{Correct Rejection}. Hence, $f$ quantifies the overall response criterion, with higher values signalling a more liberal ``old'' bias.}
    \Description{The figure presents a tree diagram explaining the Multinomial Processing Tree (MPT) model for source recognition. It begins with a test item that is either actually ``old'' or ``new.'' For old items, recognition occurs with probability d; if not recognized (1–d), the path leads to a Miss. Recognized items proceed to a Source decision with probability s. Correct source recall results in a Hit Correct outcome depending on whether the true source is ``withAI'' or ``noAI.'' If the source is not recalled (1–s), the model includes a guess step with probability beta for guessing ``withAI'' and 1–beta for guessing ``noAI,'' producing either Hit Correct or Hit Incorrect outcomes. For new items, a response of ``old'' occurs with probability f, leading to a guess step and possible False Alarm (withAI or noAI); a ``new'' response (1–f) is a Correct Rejection. Colored labels distinguish MPT model parameters (blue), incorrect outcomes (red), and correct outcomes (green). Text below the diagram describes each parameter and its interpretation in the model.}

    \label{fig:MPT}
\end{figure*}

Because joint trees of considering idea and elaboration recognition together failed to reproduce the strong within-item covariance,
we modelled \emph{idea} and \emph{elaboration} responses with two independent 2-high-threshold (2-HT) MPTs \cite{snodgrass1988pragmatics}.  
The component-wise 2-high-threshold model assigns a single probability to each latent processing step that links the unobserved memory state to the observed categorical response. \revision{Practically, this means that for a given participant and component (idea vs elaboration), the model assumes a probability of detecting seen items ($d$), a probability of remembering its source conditional on detection ($s$), a probability of guessing ``with AI'' when the source is not remembered ($\beta$) and one false alarm rate for distractors ($f$), averaged across all items of that type.} All parameters are constrained to the unit interval and are estimated hierarchically (i.e., a group mean and an across-person standard deviation). For every component, the four free parameters have the psychological interpretation shown in \cref{fig:MPT}.

All parameters were estimated hierarchically with
\texttt{TreeBUGS}~\cite{heck2018treebugs}
using 40\,000 MCMC iterations
(after 5\,000 adaptation), three parallel chains, and a thinning interval of 10. \texttt{TreeBUGS} fits MPT in a Bayesian hierarchical framework, allowing the researcher to account for repeated measures and multiple subjects, sampling the joint posterior of individual and group parameters with \textsf{JAGS} and returning full uncertainty
intervals plus posterior-predictive $p$-values for mean and covariance
structure. We used standard priors to facilitate convergence. Gelman--Rubin diagnostics were excellent ($\hat R\le 1.01$).
Posterior-predictive checks showed small mean discrepancies (T1\,obs--pred\,$<5$), but covariance misfit (T2, $p=0$) indicates residual trial-level dependence that separate trees cannot capture. \revision{This aligns with the idea that individual solutions vary in how diagnostic their style and content are for identifying the source (see Section 4.5.2). Consequently, the MPT parameters should be interpreted as summaries of average behaviour across the item set rather than precise probabilities for each individual solution.}
Parameter estimates are nevertheless robustly addressed for our preregistered H1.

\subsubsection{Idea component}

For ideas, the model estimated a hit rate of $d_{\text{idea}}=.86$ (95\% HDI $[.84,.88]$), indicating that the vast majority of studied ideas were correctly recognised as \emph{old} (\revision{i.e. ideas the participant had worked on in Phase 1}).
Conditional on a hit, the true source was recalled on $s_{\text{idea}}=.39$ of trials ($[.36,.43]$), so source memory accompanied about two out of five detections.
When the source was forgotten, participants guessed ``AI'' with probability $\beta_{\text{idea}}=.43$ ($[.39,.47]$), indicating a bias toward guessing \emph{\noAI{}} (the HDI excludes $.50$).
The false-alarm rate for lure ideas was $f_{\text{idea}}=.29$ ($[.26,.33]$), reflecting a somewhat liberal old–new criterion.

\subsubsection{Elaboration component}

Elaborations yielded the same high hit rate as ideas, $d_{\text{elab}}=.86$ ($[.84,.88]$), but with a markedly higher source-memory probability, $s_{\text{elab}}=.62$ ($[.57,.65]$), showing that source was correctly identified on well over half of detections.
When source memory failed, participants showed a bias toward AI: $\beta_{\text{elab}}=.60$ ($[.56,.64]$), with the HDI clearly above $.50$.
The false-alarm rate was again $f_{\text{elab}}=.29$ ($[.26,.33]$), closely matching the idea component.

\subsubsection{Comparison}

Posterior contrasts (Elaboration $-$ Idea) confirmed a sizeable source-memory advantage for elaborations and a stronger AI-guessing bias when memory failed:
$\Delta s = .23$ (95\% HDI $[.17,.28]$, $\Pr(\Delta s>0)=1$) and
$\Delta\beta = .18$ ($[.12,.23]$, $\Pr(\Delta\beta>0)=1$).
By contrast, sensitivity and criterion were indistinguishable across components:
$\Delta d \approx 0$ and
$\Delta f \approx 0$.

\subsubsection{Summary}
Taken together, the results reveal a clear \emph{memory advantage} for elaboration sources ($s_{\text{elab}}>s_{\text{idea}}$) and a selective \emph{AI guessing bias} on elaborations ($\beta_{\text{elab}}>.50$ and $>\beta_{\text{idea}}$), while overall recognition sensitivity and response criterion are effectively identical for ideas and elaborations. %
\revision{Accordingly, we interpret AI-related differences as reflecting changes in source monitoring and a stronger bias to attribute elaborations to AI ($\beta_{\text{elab}}$), not as changes in item recognition or discriminability (i.e., between \textit{old} and new \textit{items}, see \cref{sec:mpt}).}

\subsection{Qualitative Results}
To contextualise our findings, we examined participants' feedback, and evaluate potential patterns.

\subsubsection{Anticipatory Strategies}
In Phase 1, participants were kept unaware that they would have to remember their generated solutions in Phase 2.
To test participants' awareness of the study's true aim, at the beginning of Phase 2 participants had to guess what we will ask them during the questionnaire.
Only 18 participants ($<\pct{10}$) correctly guessed that the study was related to remembering previous ideas or solutions.
Thus, we deem it highly unlikely that participants used external support (e.g. keeping notes about Phase 1).

\subsubsection{Source Attribution Strategies}\label{sec:results_attribution_strategies}
At the end of the study, we asked participants ``Did you employ any strategies for remembering and deciding which ideas or elaborations you came up with on your own or with AI support?''.

While some participants (\pct{36.4}) did not mention any explicit strategies beyond memory to determine whether an idea or elaboration was created with or without the help of AI, others provided more insights into their thought processes.
Many (\pct{58.2}) mentioned that the writing style, such as tone, grammar, text length, or language (i.e. British vs. American English), could sometimes help to determine whether a text was written with or without the support of AI.
A few participants (\pct{11.4}) found that examining a text's content could provide useful cues. For example, they mentioned that the presence of brand names (likely the ones we used for the known-topic distractors) or highly technical terms and very creative ideas hinted at texts generated with AI-support. The participants assessed whether they could have come up with the ideas themselves or not. Two participants mentioned that they chose AI when they were unsure. 

Some participants also revealed in which cases it was particularly easy or difficult to determine the source.
Six participants explained that determining the source was easier for topics which they were already familiar with, e.g. one participant answered ``The ones related to ocean plastic I remember mostly being related to me because I am very knowledgeable about that topic, so that made those decisions easier''. Four participants mentioned that recalling their own ideas was easier. One mentioned that the interesting ideas were easier to remember. One participant commented that attributing the source for ideas was more difficult than for elaborations, while another commented that it was easier.

\subsection{Summary}
We summarise the findings as follows:

\begin{itemize}[leftmargin=*, itemsep=4pt]
    \item \textbf{AI involvement reduced item memory accuracy}, dropping from 97.6\% without AI to 87.9\% for \withAI{}/\withAI{}. 
    This corresponds to substantially lower odds of correct recognition with AI, especially at the elaboration stage (94.7\%, OR = 0.30), compared to a smaller reduction for AI-ideation (97.3\%, OR = 0.61).
    \item \textbf{Source memory is supported by \emph{consistent} workflows:}
    For remembering idea sources, accuracy was highest without any AI (92.4\%), followed by \withAI{}/\withAI{} (79.3\%), whereas it collapsed in the mixed workflows (\withAI{}/\noAI{}: 37.7\%; OR = 0.05, \noAI{}/\withAI{}: 64.0\%; OR = 0.14). %
    Remembering elaboration sources showed a similar yet less extreme pattern: \noAI{}/\noAI{} (91.5\%) and \withAI{}/\withAI{} (89.2\%) outperformed mixed cases (\withAI{}/\noAI{} 79.0\%, OR = 0.35; \noAI{}/\withAI{} 80.0\%, OR = 0.37). 
    We validated these patterns with the MPT model.
    
    \item \textbf{AI involvement reduced confidence:}
    Idea source confidence dropped from 85.2 without any AI down to 78.9 for \withAI{}/\withAI{}. 
    Costs were ca. 5-6 points when AI entered at either stage.
    Elaboration source confidence also dropped, from 86.3 without any AI to around 83 across AI-involved workflows. 
    Metacognitive sensitivity (being more confident when correct) was weak for idea attributions in mixed workflows, but strong for elaboration attributions, with the highest in \withAI{}/\withAI{}.

    \item \textbf{People overestimated their performance,}
    more strongly for ideas (by \pct{12}) than elaborations (by \pct{6}).
    Correlations between objective and perceived accuracy were weak and non-significant for ideas ($\rho = .13$, $p = .077$), but significant for elaborations ($\rho = .25$, $p < .001$).
    
\end{itemize}

\section{Discussion}
In this study, we explore how well people remember the sources of ideas and texts when working with AI, addressing our research questions: 

Our findings show negative effects of AI (RQ1): Recognition of having worked on a solution a week prior was generally high, but AI involvement reduced it. Source attribution for both ideas and elaboration texts was also highest without any use of AI. In summary, using AI reduced people's ability to remember sources and outcomes of the creative work. We term this effect the ``AI Memory Gap.''

Moreover, source attribution accuracy was lowest in mixed workflows (AI support for either ideation or elaboration), indicating that keeping workflows with AI consistent supports memory -- although this still did not reach the accuracy of not using AI (RQ2).

MPT modelling confirmed that the effects were not due to diminished sensitivity to old items. %
It confirmed weaker source memory for ideas than elaborations\revision{, likely because only the elaborations were shown verbatim, giving them richer, more distinctive, and perceptually detailed memory cues. This interpretation aligns with the SMF, which predicts that externally presented materials (such as the visible elaborations) provide more perceptual and contextual detail for memory to work with than internally retrieved content (such as the underlying ideas or thoughts), giving them a stronger basis for later source judgments. We contextualise this in \cref{sec:textcues}.} %

\revision{Our MPT results further} revealed different guessing biases when sources were forgotten -- towards self for ideas and towards AI for elaborations.
\revision{From an SMF perspective, this asymmetry might be explained by differences in the qualitative characteristics used during source monitoring. 
Ideas relate to internal cognitive operations and often lack distinctive perceptual detail; when memory is weak, participants may therefore rely on heuristic cues that make such abstract, thought-like content (i.e. underlying ideas) feel self-generated; an effect that may be amplified by having further developed these ideas during the elaboration task. 
In contrast, elaborations consist of fluent and sentence-level text-features that participants may associate with external sources, such as our AI chatbot, leading to an AI-biased guess when source memory fails.}

Finally, confidence ratings mirrored these patterns: They were highest without any AI. People overall gave rather high absolute confidence ratings, even when being wrong, and slightly overestimated their overall performance at the end. This indicates that they tended to be overconfident about their own performance (RQ1, RQ3).
\revision{At the same time, the decrease in confidence in AI-supported workflows likely reflects an implicit metacognitive sensitivity to weaker or less distinctive memory cues in these conditions. This suggests a limited awareness that source attribution is more difficult in human-AI workflows, even though confidence still exceeded actual accuracy and participants did not reliably identify when AI had been used.}

\subsection{Implications}
Following \citet{vanberkel23implications}, we identify four key implications of our work for HCI research: theory, design, community, and practice.

\subsubsection{Theory}
Studies using the \emph{Source Monitoring Framework} (SMF) \cite{mitchell2009source} are typically conducted with fixed experimental stimuli. To our knowledge, this work is the first to practically apply the SMF to interaction with generative AI.
Our results show that AI use for ideation and writing affects source monitoring, highlighting both the possibility and the importance of updating SMF theory and methodology to such human-AI interaction settings.
Concretely, the SMF posits that people attribute mental content to a source by evaluating features of the memory trace (e.g., perceptual details, spatio-temporal context, affect, and records of cognitive operations). Our findings motivate further investigation and extending these in the context of interaction with AI.

\subsubsection{Design}
These theoretical insights have clear implications for design.
AI outputs often lack the perceptual and contextual cues that the SMF identifies as essential for accurate source attribution, making it especially easy to misremember their origin.
To directly support memory in human-AI co-creative processes, interaction and UX design should embed salient cues that help users retain a ``memory of self-generation.''
For example, designs might explore using varying animations, colours, interaction modes, etc. during content generation, which are distinct between own creation vs AI use. 
\revision{More broadly, future designs could explore leveraging the concepts of UI fragmentation~\cite{Buschek2024collage} and seamful design~\cite{Inman2019seams} in context of the design dimensions for AI writing tools by \citet{Lee2024designspace} to achieve such cues.}

Alternatively, users could be supported in tracking interactions, as demonstrated by the prompt history by \citet{Hoque2024hallmark}. %
This could empower users to later verify or reconstruct their own creative process to preserve the memory of who contributed what.
In contexts where accurate source attribution is important, it could become critical to design such systems, where both the static text and process traces are part of the final output.

\subsubsection{Community}
Our work also helps the HCI community contextualize prior research on human-AI interaction, particularly around factors such as responsibility and accountability.

For example, it extends the findings of \citet{credit_he2025} by showing that disputes over credit in human-AI co-creation may not only stem from social judgments of fairness, but also from systematic memory failures that make accurate source attribution difficult, especially in mixed workflows.

The uncovered ``AI memory gap'' also provides a psychological lens for connecting with explicit source attribution effects such as the \textit{AI Ghostwriter Effect} identified by \citet{draxler24_ghostwriter}: Even when users privately feel little ownership of AI generated content, they nonetheless refrain from publicly declaring AI as the author of their creations.
Our findings work in parallel as people who genuinely misremember also fail to distinguish the origin of AI-generated ideas.
Taken together, both effects can contribute to incorrect attributions that blur the line between accidental and intentional failures to declare AI authorship.

Furthermore, our cognitive perspective adds empirical support for transparency frameworks and themes like those concerning AI disclosure obligations raised by \citet{ali24_disclosure_obligations}.
If people naturally misremember AI involvement, then voluntary disclosure is unreliable, especially after a longer time gap.
Our findings therefore suggest that disclosure obligations cannot depend only on user intent or recollection, but may require transparent, system-supported interaction logging.
Ultimately, recognizing and mitigating these memory failures is essential for designing responsible co-creation processes in human-AI interaction.

\subsubsection{Practice}
Practitioners working in environments where correct source attribution is critical (e.g., creative writing, education, medicine, and law) should be made aware that relying solely on memory can be misleading.
This is especially important in the context of our findings on perceived performance:
People working with AI tend to overestimate their ability to correctly attribute the source of their creation.
As such, end-users should be supported to critically evaluate own recollections in their collaborative human-AI work.

\subsection{Principles of Human-AI Interaction in the Context of the AI Memory Gap}
Grounded in our results and their implications, we distil a set of principles to guide the design, evaluation, and practical use of AI in human-AI interaction.

\paragraph{Principle of Explicit Source Attribution}
As AI involvement systematically impairs source memory, people should not solely (have to) rely on recollection when attributing the source of content.
As such, designers should make authorship transitions visible, evaluators should assume that memory-based self-reports are unreliable, and practitioners should adopt strategies that externalise authorship rather than depending on their own memory.
Concretely, designers could implement revision histories that highlight which passages were written by a human or suggested by an AI system. 
Evaluators such as teachers or reviewers could request that submissions include these logs alongside the final product.
Practitioners might adopt habits of annotating drafts or saving intermediate snapshots to mark when and how AI was used, in a level of detail that goes beyond merely marking text that was fully AI generated, but also to record where underlying ideas and concepts originated from.

\paragraph{Principle of Consistency}
Our findings show that consistent workflows support more accurate memory.
\revision{Users who rely on their memory, such as when systems provide limited provenance, benefit when a sequence from ideation through elaboration is completed either entirely on their own or entirely with AI.}
\revision{More consistency can also be achieved by setting ``boundaries'' and ``limiting AI to specific tasks'', although this might only apply to a subgroup of writers who want to retain authenticity, a feeling of ownership and control, or enjoyment of the craft~\cite{guo25pentoprompt}.}
\revision{In practice, this might be difficult to realise for many as AI is often integrated in flexible and opportunistic ways.
Still, the principle can be supported by keeping certain aspects of writing workflows consistent: Even if users mix AI into different stages of their process, the system itself should stay consistent.
Critically, current AI writing assistants (e.g., ChatGPT) tend to offer a ``next step''.
For example, when asked to proofread text, the AI may also suggest conceptual improvements, producing an output that mixes contributions across idea and elaboration stages.
Keeping the AI's role consistent, even as users employ it at different moments, may therefore support better memory performance.
These stable roles could be reinforced through consistent interface designs, such as placing idea-generation features in a separated UI view, while editing assistance occurs in inline interactions.}
Finally, evaluators should recognise that \revision{inconsistent} processes are more prone to memory errors.

\paragraph{Principle of Calibrated Confidence}
Evaluators should be cautious about interpreting confident statements after using AI systems as evidence of accuracy, as it does not always track accuracy, in particular for ideas in mixed workflows. %
On the flip side, users should adopt habits of double-checking attributions after having used AI in their workflows and not only trust on their own feelings. %

\paragraph{Principle of Cognitive Visibility}
While some systems make AI contributions more transparent (e.g., \citet{Hoque2024hallmark}), contributions to human thought processes are far harder to capture and later recall. %
Designers might experiment with tools that record prompts and drafts together, creating a map of thoughts alongside the evolving ideas and making AI's influence on cognitive processes more visible.
However, evaluators should acknowledge that source attributions will always be incomplete and avoid demanding certainty where it cannot exist.
Practitioners should be aware that AI may shape not only their writing but also their thinking, often in ways that are difficult to recognize, even for the authors themselves.

\paragraph{\revision{Contextualisation with other principles}}
\revision{We contextualise these principles with two widely recognised sets of human-AI interaction guidelines in the literature.
First, the 18 guidelines by \citet{Amershi2019guidelines} are structured along a timeline from ``initially'' to ``during interaction'' to ``over time''. While \textit{human} memory is not explicitly considered, it might best fit with design choices in line with guideline 12, which says that \textit{systems} should ''Remember recent interactions''. Moreover, our work motivates an additional category: ``after active use''.  
Second, we relate our work to the six design principles for generative AI applications by \citet{Weisz2024principles}. While these do not mention memory, our insights could extend their principles of ``Design for Mental Models'' or ``Design for Imperfection'', if the latter is conceptually expanded from meaning system imperfections to also covering imperfect user memory.}

\subsection{Effective Memory Engagement}
\label{sec:eme}

Here, we dissect our results in light of potential concerns that correct answers could arise without genuine memory engagement.
Asking about presented information inadvertently comes with possibilities for answering (correctly) even without actually remembering -- in our case by (1) guessing or (2) inferring the answer from cues in the presented text (e.g. ``AI-typical'' phrasing). We discuss both these ``alternative pathways'' below.

\subsubsection{Memory vs Guessing}
Participants performed well above the guessing baseline of 50\% in all cases, except for idea source attribution in the \withAI{}/\noAI{} workflow, where they were systematically worse. 
Moreover, confidence tracked accuracy in most cases, whereas guessing typically produces low subjective certainty~\cite{siedlecka2021confidence}.
Thus, our results clearly show that people did not attempt to solve the tasks mainly by guessing.

\subsubsection{Memory vs Text Cues}
\label{sec:textcues}
Participants only attributed sources after identifying an item as seen. 
Already for this first step, \textit{any} use of AI had a significantly negative influence, indicating that memory formation was affected in Phase 1.

Moreover, source memory accuracies for elaborations (\pct{80} - \pct{90}, cf. \cref{tab:emm_elab}) are substantially higher than accuracies reported in studies where humans try to identify AI-generated text (\pct{50} - \pct{55}, cf. \cite{clark2021thatshumangoldevaluating, Jakesch_heuristics_2023, boutadjine2025human}).
This strongly suggests that participants' performance was driven by a genuine memory effect rather than mere reliance on heuristic detection cues, such as those \revision{described} in \cref{sec:results_attribution_strategies}.

To dissect this further, the MPT model is specifically designed to isolate item memory, source memory, and guessing. It showed that, after correctly detecting an item as seen, participants recalled the true source without guessing in 39\% of cases for ideas and 62\% for elaborations. While text cues -- which are likely more directly indicative for elaborations -- might contribute to the difference between these two values, this does not take away from our result of the ``AI memory gap'': 
That is, isolated from guessing in this way, and combined with the results on accuracy, we see that even with potential text cues, source memory for both idea and elaboration was far from perfect -- and significantly worse with AI.

\subsection{Limitations, Generalisability, and Future work}
\revision{Here, we discuss limitations, generalisability of our findings, and future research directions.}

\subsubsection{\revision{Writing Tasks}}
We only examined the impact of AI usage during ideation and elaboration tasks as part of the writing process. 
We opted for those tasks as they represent fundamental stages that precede and influence many other creative processes.
\revision{However, many real-world uses of LLMs involve different cognitive demands and output structures. Future work could therefore examine whether similar AI-induced source confusion occurs in tasks such as summarisation, paraphrasing, document search, information extraction, or multimodal analysis.}

\revision{We hypothesise that these tasks differ in ways that may widen or close the AI memory gap. For instance, summarisation and paraphrasing often build directly on previous text and may blur the boundary between self- and AI-generated content even further, especially in mixed workflows. Remembering whether one conducted document or information search may also be more difficult: AI-mediated search removes users from the step-by-step exploration that normally creates distinctive memory cues (e.g., navigating papers or websites), and AI systems frequently pre-extract relevant information directly in the chat window, further reducing contextual markers. Conversely, tasks that require substantial transformation or personal reasoning may preserve stronger cues of self-authorship.}

\revision{Related,} whether our findings apply to domains other than writing, such as coding, image and video creation, or design, needs to be explored in future work. 
\revision{In addition}, our experimental task mimicked a co-creative workflow for all AI conditions, where the human always made the final decision.
Situations in which AI makes executive decisions on its own remain open for investigation.
Based on our results, we hypothesise that this would further reduce people's ability to remember parts of the creative process.
\revision{A needed next step is to identify which task characteristics and interface cues most strongly shape the AI memory gap, with a study in the field and with full writing workflows.}

\subsubsection{\revision{AI Interactions}}
Our study focused on a small subset of possible AI interactions, specifically using an LLM (GPT-4.1-mini) through a chatbot interface.
Using different models, interaction paradigms (e.g., inline suggestions), or interfaces could lead to different outcomes.
\revision{Moreover, participants completed the Phase 2 memory tasks without access to any external aids, such as prompt histories or chat logs.}
\revision{This was intentional, as our aim was to measure source memory ability in isolation.} 
\revision{However, in practice, users may have access to persistent chat histories, document revision trails, or retrieval features. Such aids do not necessarily improve memory directly, but they do preserve interaction traces that users can revisit, which may change how source attribution is solved in a specific practical context.}

\subsubsection{\revision{Ecological Factors}}
In our study, people created ideas in one session, whereas \revision{typical} creative workflows often stretch over multiple sessions. 
We also tested people's source memory after a fixed time gap of approximately one week, and it remains unclear how shorter or longer retention intervals might influence memory performance in this context.
\revision{Moreover, individuals' motivation and effort to recall AI contributions may vary in different real-world settings, for example when people feel pressure to justify their choices in a group setting or when they are simply exploring ideas informally for personal inspiration.
Future work could also examine memory performance when participants are informed in advance that they will later be asked to report their use of AI assistance.}

\section{Conclusion}
Our study shows an ``AI memory gap'': Working with AI systematically impairs people's ability to accurately recall the source of their creative contributions. 
Memory for authorship was strongest in fully human workflows and weakest in mixed human–AI settings, where mismatched sources disrupted attribution the most.
While participants often felt confident in their performance, this confidence exceeded their actual accuracy. 
Our findings underline the risks of misattribution in human–AI collaboration and call for system designs that make authorship more transparent, as well as further research into how different AI roles, modalities, and interaction settings shape memory and responsibility in creative work.

\begin{acks}
This project is funded by the Deutsche Forschungsgemeinschaft (DFG, German Research Foundation) -- 525037874. This work relates to the upcoming ERC project AmplifAI (grant agreement No. 101217557). Daniela Fernandes is funded by the Finnish Doctoral Program Network in Artificial Intelligence, AI-DOC (decision number VN/3137/2024-OKM-6).
\end{acks}

\bibliographystyle{ACM-Reference-Format}
\bibliography{bibliography}

\appendix

\section{Appendix}
\subsection{Problem Statements}
\label{sec:appendix_statements}

\begin{enumerate}
    \item Come up with simple actions or programs a company can take to help workers spot and avoid scam emails.
    \item Think of features that could help blind or low-vision people use a city’s bus/train app more easily.
    \item What rules or actions could social media platforms or governments take to slow down or stop fake videos from being shared?
    \item What practical things can a family doctor do to help older adults remember and stick to their medicine routine?
    \item Come up with ways students can fairly distribute work for an essay in a group project.
    \item Oceans are heavily polluted with plastic waste. What practical approaches could help reduce plastic pollution in the oceans?
    \item Many countries face demographic challenges with a growing elderly population and fewer young workers. What approaches could help ensure the sustainability of pension funds?
    \item Climate change is causing more frequent and severe weather events. What innovative ideas could help communities better prepare for and adapt to these changes?
    \item With the rise of remote work, what measures can companies take to maintain employee engagement and productivity outside a traditional office?
    \item Come up with practical solutions on how to stay cool during hot summer days.
\end{enumerate}

The known-topic and unknown-topic distractors can be found in our project repository (\url{https://osf.io/dvkj9/}).

\subsection{Questionnaire Questions}
\label{sec:appendix_questionnaire}
\begin{enumerate}
    \item How many of the ideas (in percent) do you think you assigned correctly? (0-100)
    
    \item Do you think you assigned more ideas incorrectly to AI or to yourself?
    \begin{itemize}
        \item More incorrectly assigned to AI
        \item More incorrectly assigned to myself
        \item About equally to both
    \end{itemize}
    
    \item How many of the solution texts (in percent) do you think you assigned correctly? (0-100)
    
    \item Do you think you assigned more solution texts incorrectly to AI or to yourself?
    \begin{itemize}
        \item More incorrectly assigned to AI
        \item More incorrectly assigned to myself
        \item About equally to both
    \end{itemize}
    
    \item Did you employ any strategies for remembering and deciding which ideas or elaborations you came up with on your own or with AI support?
    
    \item Gender:
    \begin{itemize}
        \item Male
        \item Female
        \item Non-binary
        \item Prefer not to say
        \item Prefer to self describe: \underline{\hspace{2cm}}
    \end{itemize}
    
    \item English Proficiency:
    \begin{itemize}
        \item Native speaker
        \item Fluent (C2)
        \item Advanced (C1)
        \item Upper-intermediate (B2)
        \item Intermediate (B1)
        \item Pre-intermediate (A2)
        \item Beginner (A1)
    \end{itemize}
    
    \item Highest Degree:
    \begin{itemize}
        \item PhD/Doctorate
        \item Master's degree
        \item Bachelor's degree
        \item Associate degree
        \item High school diploma
        \item Some college (no degree)
        \item Other
    \end{itemize}
    
    \item How often do you use AI tools (e.g., ChatGPT, Claude, Copilot)?
    \begin{itemize}
        \item Daily
        \item Several times a week
        \item About once a week
        \item A few times a month
        \item Rarely
        \item Never
    \end{itemize}
\end{enumerate}

\subsection{Misc}
\label{sec:appendix_misc}

\newcolumntype{L}[1]{>{\raggedright\arraybackslash}p{#1}}

\begin{table*}[h]
  \centering
  \begin{tabularx}{\textwidth}{@{}L{2cm}X@{}}
    \toprule
    \textbf{Section} & \textbf{Deviation} \\
    \midrule
    Participant recruitment &
    We recruited participants until reaching a sample of 200 valid and complete responses \textit{in Phase 1}. 
    To replace the drop-outs, we started 5 sequential recruitment rounds.
    Due to time-constraints we decided against starting another round because each round would delay data collection by another week.
    This resulted in the final sample of \nMain{} people who successfully completed both Phases of the study.\\
    \midrule
    Condition labels & We exchanged the condition label as used in this paper, using ``withAI'' instead of ``AI'', to clarify that ideas and elaborations were co-created by the participants together with AI, rather than generated automatically by AI alone. The label for the condition without any AI involvement remained ``noAI''. \\
    \bottomrule
  \end{tabularx}
  \caption{Rationale for deviations from the pre-registration. The pre-registration can be accessed at: \url{https://aspredicted.org/g78p-w7db.pdf}}
  \Description{The table contains two rows rationale for deviations from the pre-registration.}
  \label{tab:prereg_deviation}
\end{table*}

\end{document}